\documentclass[apjl]{emulateapj}
\submitted{}
\usepackage{graphicx}
\usepackage {natbib}
\usepackage{epstopdf}

\newcommand{\secpoint}{\mbox{$''\mskip-7.6mu.\,$}}

\shorttitle{LBG Escape Fraction at $z\sim3$}
\shortauthors{Siana et al.}

\begin{document}

\title{A Deep Hubble and Keck Search for Definitive Identification of Lyman Continuum Emitters at $z\sim3.1^{\dagger}$ $^*$}

\author{\sc Brian Siana\altaffilmark{1}, Alice E. Shapley\altaffilmark{2}, Kristin R. Kulas\altaffilmark{2, 3}, Daniel B. Nestor\altaffilmark{2}, Charles C. Steidel\altaffilmark{4}, Harry I. Teplitz\altaffilmark{5}, Anahita Alavi\altaffilmark{1}, Thomas M. Brown\altaffilmark{6}, Christopher J. Conselice\altaffilmark{7}, Henry C. Ferguson\altaffilmark{6},  Mark Dickinson\altaffilmark{8}, Mauro Giavalisco\altaffilmark{9}, James W. Colbert\altaffilmark{5}, Carrie R. Bridge\altaffilmark{4}, Jonathan P. Gardner\altaffilmark{10}, Duilia F. de Mello\altaffilmark{11}}

\altaffiltext{1}{Department of Physics and Astronomy, University of California Riverside, Riverside, CA 92521}

\altaffiltext{2}{Department of Astronomy, University of California Los Angeles, Los Angeles, CA 90095}

\altaffiltext{3}{Physics Department, Santa Clara University, 500 El Camino Real, Santa Clara, CA 95053}

\altaffiltext{4}{California Institute of Technology, MS 249-17, Pasadena, CA 91125}

\altaffiltext{5}{Infrared Processing and Analysis Center, California Institute of Technology, 220-6, Pasadena, CA 91125}

\altaffiltext{6}{Space Telescope Science Institute, 3700 San Martin Drive, Baltimore, MD 21218}


\altaffiltext{7}{University of Nottingham, Nottingham, NG7 2RD, UK}

\altaffiltext{8}{National Optical Astronomy Observatory, 950 N. Cherry Ave., Tucson, AZ 85719}

\altaffiltext{9}{University of Massachusetts, Department of Astronomy, Amherst, MA, 01003}

\altaffiltext{10}{Astrophysics Science Division, Observational Cosmology Laboratory, Code 665, Goddard Space Flight Center, Greenbelt, MD 20771}

\altaffiltext{11}{Department of Physics, Catholic University of America, 620 Michigan Avenue, Washington DC 20064}

\altaffiltext{${\dagger}$}{Based on observations made with the NASA/ESA Hubble Space Telescope, obtained at the Space Telescope Science Institute, which is operated by the Association of Universities for Research in Astronomy, Inc., under NASA contract NAS 5-26555.  These observations are associated with program 10405, 11636, 11735.}

\altaffiltext{$*$}{Some of the data presented herein were obtained at the W.M. Keck Observatory, which is operated as a scientific partnership among the California Institute of Technology, the University of California and the National Aeronautics and Space Administration. The Observatory was made possible by the generous financial support of the W.M. Keck Foundation.}

\begin{abstract}

Narrow-band imaging of the rest-frame Lyman continuum (LyC) of galaxies at $z\sim3.1$ has produced a large number of candidate LyC-emitting galaxies. These samples are contaminated by galaxies at lower redshift. To better understand LyC escape, we need an uncontaminated sample of galaxies that emit strongly in the LyC. Here we present deep {\it Hubble} imaging of five bright galaxies at $z\sim3.1$ that had previously been identified as candidate LyC-emitters with ground-based images. The WFC3 F336W images probe the LyC of galaxies at $z>3.06$ and provide an order-of-magnitude increase in spatial resolution over ground-based imaging. The non-ionizing UV images often show multiple galaxies (or components) within $\sim1''$ of the candidate LyC emission seen from the ground. In each case, only one of the components is emitting light in the F336W filter, which would indicate LyC escape if that component is at $z>3.06$. We use Keck/NIRSPEC near-IR spectroscopy to measure redshifts of these components to distinguish LyC-emitters from foreground contamination. We find that two candidates are low redshift contaminants, one candidate had a previously misidentified redshift, and the other two cannot be confirmed as LyC-emitters. The level of contamination is consistent with previous estimates. For the galaxies with $z>3.06$, we derive strong $1\sigma$ limits on the relative escape fraction between 0.07 and 0.09. We still do not have a sample of definitive LyC-emitters, and a much larger study of low luminosity galaxies is required. The combination of high resolution imaging and deep spectroscopy is critical for distinguishing LyC-emitters from foreground contaminants.  

\end{abstract}

\keywords{galaxies: high-redshift, galaxies: starburst, galaxies: intergalactic medium, ultraviolet: galaxies}

\section{Introduction}

Surveys are now identifying large numbers of galaxies at the reionization epoch ($z\sim7-8$). Consequently, there has been much interest in how those galaxies reionized the hydrogen in the intergalactic medium (IGM).  Several studies have determined that the number density of AGNs is not high enough to provide the required ionizing background \citep{inoue06,siana08,willott10,masters12}, though some have suggested that faint active galactic nuclei (AGNs) may contribute significantly \citep{glikman11,fontanot12}. Because massive stars are the only other significant source of ionizing photons, they are believed to be responsible for the reionization of intergalactic hydrogen.  

Whether or not stars reionized the intergalactic medium and provided the ionizing background for 1-2 Gyr thereafter is contingent upon many uncertain parameters:  the {\it total} star formation rate density (SFRD) including the significant contribution from sources beyond our current detection limits; the intrinsic ionizing spectrum of the galaxies; the fraction of the ionizing radiation produced by these galaxies that escapes into the intergalactic medium; and the ``clumping'' of the hydrogen being ionized in the IGM at any epoch (and thus the recombination rate).  

Many efforts, both theoretical and observational, are being made to determine the amount of undetected star formation \citep{alavi14} and the clumping factor \citep{pawlik09, finlator12}.  However, little is understood about the ``escape fraction'', $f_{esc}$, of ionizing radiation. Young stars form in dense molecular clouds. In order for a significant fraction of ionizing radiation to escape into the IGM, lines of sight (LOSs) with low neutral hydrogen column-density must be cleared out or the stars must migrate out of the gas. Either process must occur within the short lifetime ($<10$ Myr) of the O-stars that produce the ionizing radiation.  

It is not understood how the stars can be exposed on such short timescales, though several mechanisms have been proposed. These include supernovae winds and their effects on the H{\sc i} distribution in both the interstellar medium (ISM) and the circumgalactic medium (CGM) \citep{dove00, clarke02, fujita03,razoumov06}, interactions/mergers with other galaxies \citep{gnedin08a}, and runaway massive stars \citep{conroy12}.  Each of these proposed mechanisms has observational signatures that would allow us to test which processes are responsible for high escape fractions and, thus, the reionization of the universe.  

Regardless of the precise mechanism, the first goal is to measure the average escape fraction and any dependencies it might have on galaxy properties or environment. Over the last 15 years, many attempts have been made to directly detect escaping Lyman continuum (LyC) radiation from galaxies of varying types and at various redshifts \citep{leitherer95,deharveng01,steidel01,giallongo02,malkan03,siana07,grimes07,cowie09,grimes09,iwata09, bridge10,siana10, vanzella10, nestor11, leitet11, boutsia11, vanzella12, nestor13, mostardi13, leitet13, borthakur14}.  These studies must be conducted at $z\lesssim 3.5$ because the IGM becomes optically thick to ionizing photons at higher redshift \citep{inoue08, prochaska09}. Also, galaxies in the nearby universe must be receding at sufficient velocities such that a significant fraction of the ionizing spectrum is redshifted to wavelengths greater than 912 \AA\ to avoid absorption by Galactic H {\sc i} \citep{hurwitz97}. The most sensitive limits have been obtained at $z\sim1$ with ultraviolet imaging or spectroscopy with the $Hubble$ Space Telescope ($HST$) and $GALEX$ \citep{malkan03,siana07,cowie09, bridge10,siana10} and at $z\sim3$ with optical data from the ground \citep{shapley06,iwata09,nestor11,nestor13,mostardi13}.  The results of the studies at $z\sim1$ and $z\sim3$ are quite different.  At $z\sim1$, there are no detections from a sample of 68 galaxies \citep{siana07,siana10, bridge10} and a null result from a stack of 626 galaxies \citep{cowie09}.  At $z\sim3$ however, roughly 10\% of galaxies have been detected with large ionizing emissivities \citep{iwata09, nestor11, nestor13}.  Because the limits at $z\sim1$ are typically far deeper than the surveys at $z\sim3$, the large number of detections at $z\sim3$ suggests an average escape fraction that is significantly higher at $z\sim3$ \citep{siana10}.

There are several possible explanations for the seemingly discrepant results at $z\sim1$ and $z\sim3$.  First, the two samples were selected in different ways.  In particular, most of the galaxies in the $z\sim3$ samples were selected as Ly$\alpha$-emitters. Because both the Ly$\alpha$ and LyC escape are dependent upon the distribution of neutral Hydrogen, it is certainly possible that a Ly$\alpha$ selection would also tend to find LyC emitters.  However, \citet{nestor11} saw {\it lower} Ly$\alpha$ fluxes for galaxies with LyC detections, suggesting that Ly$\alpha$ emission is not necessarily positively correlated with high LyC escape fractions. Second, the role of environment may be important. The large samples at $z\sim3$ have been selected within large over densities to increase observing efficiency, but the increased star formation in these dense environments might affect the local ionizing background and the neutral hydrogen columns in the circumgalactic medium. Finally, the $z\sim1$ and $z\sim3$ samples may be in different phases of their evolution.  The samples have similar SFRs and stellar masses \citep{siana10}.  However, because the $z\sim1$ samples were selected to be very luminous, they lie well above the star-forming main sequence at $z\sim1.3$ \citep{whitaker12}, and are likely in a transient starburst phase.  It is possible that the beginning stages of such a starburst may not be conducive to high LyC escape fractions (for example, because gas fractions are particularly high or because many are merging systems). Finally, it may be possible that other galaxy properties that have not been measured well -- their sizes or circumgalactic medium, for example -- can influence the LyC escape fraction. 

Many uncertainties remain in understanding ionizing photon escape and the differences in the $z\sim1$ and $z\sim3$ studies. Foremost among these is the possibility of foreground contamination of the $z\sim3$ LyC detected galaxies.  In some cases, it is possible that a faint star-forming galaxy lies along the line of sight to the target galaxy, but at significantly lower redshift.  In this case a candidate detection in the rest-frame LyC of the background galaxy could simply be non-ionizing emission from the foreground galaxy.  At $z\sim1$, the line of sight is much smaller and the flux densities required to probe significant escape fractions are not as low.  Therefore, there are far fewer foreground objects along the line of sight.  In addition, the $z\sim1$ studies have been done with the {\it Hubble} Space Telescope and its high spatial resolution allows for easier identification of foreground objects at small angular separation (or impact parameter). 

Some studies have used deep U band (which probes below the Lyman limit above $z\gtrsim3$) number counts to determine the probability of foreground contamination at $z\sim3$ \citep{siana07, vanzella10, nestor11, mostardi13} and find that a significant fraction of LyC detections are in fact attributed to foreground contamination. This contamination fraction is a strong function of the depth of the LyC images \citep[since the number density of foreground objects rises steeply at fainter magnitudes, e.g. ][]{alavi14}, as well as the FWHM of the point-spread function (as the PSF widens, contaminants at larger impact parameters cannot be recognized).  

These statistical corrections for foreground contamination are useful when determining what fraction of galaxies have detectable LyC escape, or for constraining the total ionizing emissivity of galaxies. However, it is useful to know for certain {\it which} galaxies truly have high escape fractions, as they need to be studied in greater detail to understand the mechanisms for escape.

There are two ways to mitigate the foreground contamination problem.  First, deep spectroscopy of candidate LyC emitters may provide evidence of the foreground object if emission lines are detected at other redshifts. However, the foreground object may lack strong emission lines or the lines may be shifted out of the wavelength range covered by the spectrum. Other evidence for a foreground galaxy may be seen in absorption.  Any foreground contaminant would lie at a very small impact parameter ($<10$ kpc), and \citet{steidel10} have shown that galaxies generally exhibit strong metal-line absorption at these distances. A high $S/N$ detection of the continuum of the background galaxy is required to identify foreground absorption and is very difficult to obtain for galaxies fainter than $R=24.5$.  

Another way to minimize the frequency of foreground contamination is to increase the spatial resolution of the LyC imaging, as the contamination rate is proportional to the area of the seeing disk. If one resolves the background galaxies (with half-light radii of $\sim$ 0\secpoint2) instead of spreading the light over the typical seeing disc size ($0\secpoint6-1\secpoint0$), the contamination rate goes down by a factor of $\sim9-25$ and becomes almost negligible \citep{vanzella12}.  

In addition to simply removing the possibility of foreground contamination, higher resolution images may allow us to resolve the LyC-emitting regions of galaxies. If SNe winds clear low column density sightlines, the LyC may escape primarily from the regions with highest star formation surface density.  However, if galaxy interactions or runaway stars are the primary mechanism, the LyC emission will be less concentrated than the non-ionizing UV continuum. 

We have set out to determine definitively whether candidate LyC emitters have foreground contamination or indeed have high ionizing emissivities and, if the latter, which mechanisms allow high escape fractions. To do this we have followed up some of the brightest LyC-emitting candidates selected from ground-based observations by \citet{nestor11} in SSA22, a field with a large overdensity of galaxies at $z=3.09$ \citep{steidel98}.  Because of the large number of galaxies at the same redshift, this field has been used to efficiently find candidate LyC emitters with a narrow-band filter just below the observed wavelength of the Lyman limit \citep{iwata09, nestor11, nestor13}. 

First, we obtained deep $HST$ imaging of the rest-frame LyC of these galaxies to determine from which galaxy (or regions of the galaxy) the candidate LyC is escaping.  Second, we obtained deep Keck near-IR spectroscopy to detect the rest-frame optical emission lines (primarily [O {\sc iii}] $\lambda$5007) at high spatial resolution ($\sim$ 0\secpoint5) from the regions emitting apparent LyC emission to verify the redshifts of the sources. The {\it Hubble} imaging allows us to identify possible contaminating faint foreground galaxies, and the Keck spectroscopy allows us to identify the redshifts of all candidate LyC emitters.  Together, these observations allow us to detect foreground contaminants and, therefore, identify the true LyC emitters.

In Section \ref{sect:hst}, we discuss the {\it Hubble} observations.  In Section \ref{nirspec:obs}, we discuss the Keck near-IR spectroscopic observations.  In Section \ref{sect:results}, we discuss each of the galaxies individually.  Finally, in Section \ref{sect:discussion}, we put the findings in context with other recent work.

\section{HST Observations and Data Reduction}
\label{sect:hst}

\subsection{Target Selection}

\begin{center}
\begin{figure*}
\epsscale{1.15}
\plotone{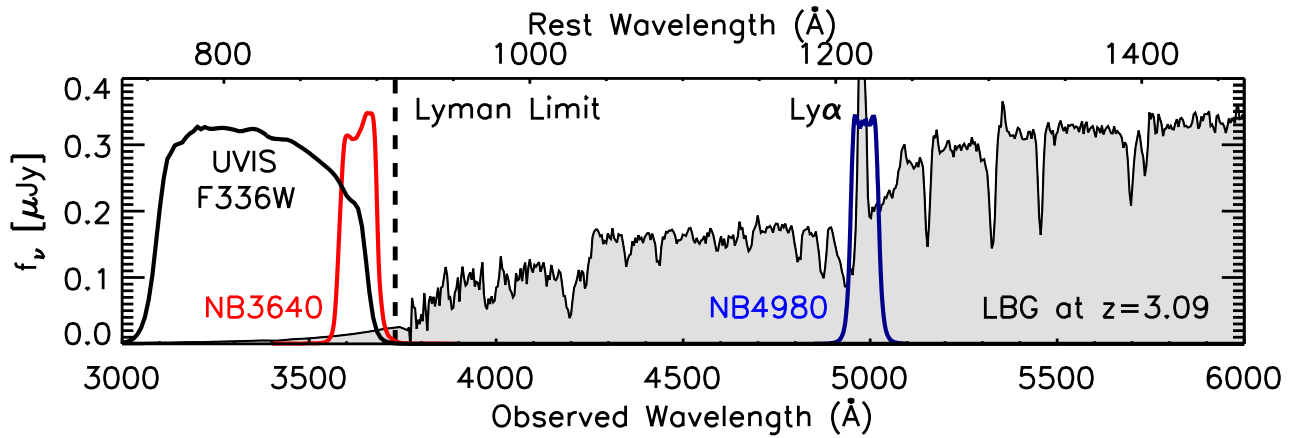}
\caption{The shaded region shows the composite spectrum of \citet{shapley03} of $z\sim3$ LBGs, shifted to the redshift of the over density in SSA22, $z=3.09$. The spectrum below rest-frame 920 \AA\ is just an extrapolation assuming a constant value in $f_{\nu}$ and multiplying by the average transmission of the intergalactic medium at  $z=3.09$, exp($-\tau_{IGM}(\lambda)$).  Also plotted are the Keck/LRIS transmission curves for the NB4980 filter (used to find Ly$\alpha$-emitters at the redshift of the over density) and the NB3640 filter (used to probe the LyC photons from galaxies at $z>3.09$). Finally, we plot the total system throughput for the WFC3/UVIS F336W filter - also used to detect LyC photons from galaxies at $z>3.09$. F336W is a broader and bluer filter than NB3640.  Depending upon  the line-of-sight IGM opacity, the F336W filter can detect more LyC photons than the NB3640.  However, the {\it average} flux density across the filter will be much lower. }
\label{fig:spec_filt}
\end{figure*}
\end{center}

Our target galaxies are all bright, $L \gtrsim 0.5L^*$ LBGs at $z>3.06$ in the SSA22 field and were originally catalogued in \citet{steidel03}. The LyC region of these galaxies at $z>3.06$ was imaged with a blue narrow-band filter with the Low Resolution Imaging Spectrograph (LRIS) on Keck I (NB3640, see Figure \ref{fig:spec_filt} for the narrow-band filter curves) by \citet{nestor11}. The galaxies were selected by their Ly$\alpha$ emission at $z\sim3.09$ with a narrow-band filter (NB4980) or via a Lyman break selection using broad-band filters \citep{steidel03}. The {\it Hubble} WFC3/UVIS field-of-view ($2.7' \times 2.7'$) is considerably smaller than that of Keck/LRIS ($5.5' \times 7.6'$) and a very deep exposure is required to detect the rest-frame LyC of these galaxies with {\it Hubble}.  Therefore, we chose to obtain a single, deep $HST$ LyC image near the NW corner of the LRIS image that has several bright candidate LyC emitters.  The bright galaxies lying in the single WFC3/UVIS pointing are MD32, C49, MD46 and D17.  There is an additional candidate, aug96M16, that does not lie in this {\it Hubble} field, but we obtained Keck/NIRSPEC spectroscopy of that source as well (see Section \ref{nirspec:obs}).  

\subsection{Hubble Observations}

We were allocated a total of 39 orbits in Cycle-17 (PID: 11636).  All primary observations are summarized in Table \ref{tab:hst_obs}.  Most (32) of the orbits were dedicated to deep imaging with the F336W filter, which probes the rest-frame LyC of galaxies at $z>3.06$ (see filter curve in Figure \ref{fig:spec_filt}).  The required exposure time in the F336W filter is large, as the expected ionizing continuum is much lower than the non-ionizing UV continuum.  Below $\sim 4000$ \AA, read noise is the dominant source of noise in UVIS imaging.  Therefore, the F336W exposure times were long (half-orbit in length, 1325s) to minimize the number of read outs. The 32 orbits were split into two-orbit visits.  In each visit, a standard four point dither (WFC3-UVIS-DITHER-BOX) was used to achieve sub-pixel sampling of the point-spread function (PSF).  Each visit performed the same dither pattern, but at slightly different central pointings spread over $\sim3''$.  

In addition to the F336W images, we also obtained optical imaging (F606W, F814W, which probe the non-ionizing rest-frame UV continuum of the targets) with the Wide Field Camera on the Advanced Camera for Surveys (ACS/WFC) and near-infrared imaging (F110W, F160W, which probe the rest-frame optical continuum of the targets) with the WFC3 infrared channel (WFC3/IR). The field was imaged for one orbit in each of these additional filters with a standard four-point sub-pixel dither pattern (ACS-WFC-DITHER-BOX, WFC3-IR-DITHER-LINE).  Two pointings were required with WFC3/IR to image all of the targets in the field because the IR channel has a smaller field-of-view than that of the UV channel.  

In addition, there is a 3-orbit ACS/WFC F814W image from a Cycle-13 program (PID 10405, PI: Scott Chapman) which covers most of our field and which we incorporated into our data.

Finally, one of the objects in this study (aug96M16) lies far outside the UVIS field-of-view.  There is existing ACS F814W imaging (3 orbit depth, PID 10405) and F160W imaging (one orbit depth, PID 11735) that we reduced and shifted to the same astrometric system as the Keck/LRIS images. 

\begin{center}
\begin{deluxetable}{lrrrr}
\tablecaption{HST Observation Summary}
\tablehead{\colhead{Camera} & \colhead{Filter} & \colhead{Orbits} & \colhead{Exptime} & \colhead{Depth\tablenotemark{a}}}
\startdata
WFC3/UVIS                             & F336W & 32 & 84800 & 29.27 \\
ACS/WFC                                & F606W & 1   & 2140   & 27.69 \\
ACS/WFC\tablenotemark{b}   & F814W & 1   & 2280   & 27.28 \\
ACS/WFC\tablenotemark{b}   & F814W & 3   & 6144   &   28.05 \\
WFC3/IR (East)                                 & F110W & 1   & 2616   & 28.07 \\
WFC3/IR (West)			& F110W & 1 & 2616 & 27.74 \\
WFC3/IR (East)                                & F160W & 1   & 2616    & 27.42 \\
WFC3/IR (West)			& F160W & 1 & 2616 & 27.19 \\
WFC3/IR (aug96M16) \tablenotemark{c}               & F160W & 1 & 2612 & 27.43 
\enddata
\label{tab:hst_obs}
\tablenotetext{a}{$3\sigma$ depth in a 0.5$''$ diameter aperture}
\tablenotetext{b}{Most of the area was observed previously with ACS/WFC F814W to three orbit depth in Cycle 13 (PID: 10405; PI: S. Chapman).}
\tablenotetext{c}{F160W image obtained in Cycle 17 (PID: 11735; PI: F. Mannucci).}
\end{deluxetable}
\end{center}

\subsection{Data Reduction and Analysis}

The {\it HST} data were reduced using multidrizzle, the standard PyRAF routines provided by the Space Telescope Science Institute (STScI). The flattened frames were cleaned of cosmic rays and stacked onto images with matching WCS and aligned within a fraction of a pixel.  Because sub-pixel dither patterns were used, we sample the PSF on a finer scale than the native pixel scales of 0\secpoint04 for WFC3/UVIS, 0\secpoint05 for ACS/WFC and 0\secpoint13 for WFC3/IR. Therefore we drizzled the data onto images with pixel sizes of 0\secpoint02, 0\secpoint04, and 0\secpoint08 for WFC/UVIS, ACS/WFC, WFC3/IR, respectively. Using stars in the field, we measure PSF FWHMs of 0\secpoint07, 0\secpoint09, and 0\secpoint18 in the WFC3/UVIS, ACS/WFC, and WFC/IR channels.

The output variance maps from multidrizzle were used to determine the depths of the images.  These depths agree well with the pixel-to-pixel variations in the images once they are corrected for correlated noise via the methods in \citet{casertano00}.  The $3\sigma$ depths are listed in Table \ref{tab:hst_obs}.

\begin{figure*}
\epsscale{1.1}
\plotone{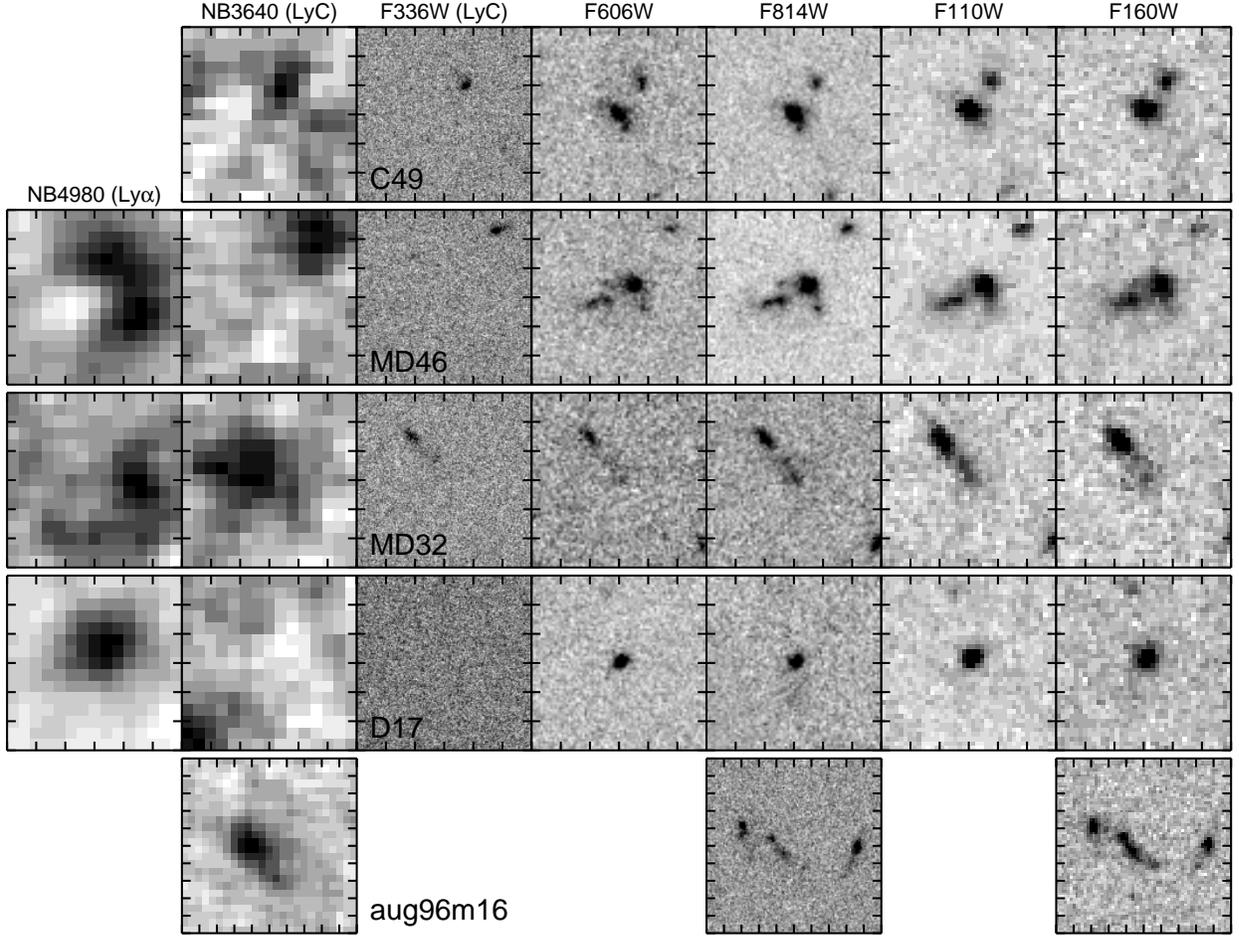}
\caption{$Hubble$ imaging of the five bright LBGs with Keck NB3640 detections from \citet{nestor11} (candidate LyC-emitters).  The stamps are 3$''$ on a side, with tick marks at 0\secpoint5 intervals (the aug96m16 stamp at the bottom is 5$''$ on a side). In C49, MD46, and MD32, the F336W (possible LyC) flux is emitted from only one of multiple clumps. The D17 detection is the low surface brightness detection in the top left of the NB3640 image.  The $HST$ imaging is not as sensitive to low surface brightness emission so it can not definitively detect it.  In the aug96m16 stamp the central galaxy has strong emission in the NB3640 filter but this galaxy lies outside of the footprint of the UVIS pointing.}
\label{fig:stamps}
\end{figure*}

We combined the F110W and F160W images (weighting by the inverse variance in each image pixel) to produce a deep detection image.  We ran SourceExtractor \citep{bertin96} to identify sources above the background fluctuations.  The isophotes from the deep detection image were then mapped onto the finer pixel scales of the WFC3/UVIS and ACS/WFC images.  Because the images were drizzled to pixel scales of 0\secpoint02, 0\secpoint04, and 0\secpoint08 (in WFC3/UVIS, ACS/WFC, and WFC3/IR, respectively), each pixel in the WFC3/IR image has either four ACS/WFC pixels or 16 WFC3/UVIS pixels in it. Therefore, the isophotes from the detection image can be mapped directly to the finer pixel images.  Finally, we used our own code to determine isophotal fluxes in the same isophotes in all filters.  The images of the targets are displayed in Figure \ref{fig:stamps} and fluxes are listed in Table \ref{tab:props}.  

In addition to the fluxes, the estimated relative escape fractions (or the $1\sigma$ limits) are listed as well.  The relative escape fraction, $f_{esc,rel}^{LyC}$ is defined as in \citet{siana07}:

\begin{equation}
f_{esc,rel}^{LyC} = \frac{(f_{1500}/f_{LyC})_{stel}}{(f_{1500}/f_{LyC})_{obs}} \ \textrm{exp}(\tau_{HI,IGM})
\label{eqn:fesc}
\end{equation}

\noindent where $f_{1500}$ and $f_{LyC}$ are the flux densities, in $f_{\nu}$, at 1500 \AA\ (F814W fluxes) and in the LyC (F336W fluxes, $\sim820$ \AA\ for these observations), respectively.  The LyC flux density is assumed to be flat (constant) in $f_{\nu}$ before IGM attenuation and the intrinsic amplitude of the Lyman break in the stellar SED is assumed to be $(f_{1500}/f_{LyC})_{stel} = 6.0$ \citep{siana07}.  We note that this ratio can be affected by short term variations in the star formation history.  Specifically, after a rise in the star formation rate, the LyC will be significantly enhanced compared to the 1500 \AA\ continuum.  Conversely, shortly after a downturn in the star formation rate, the LyC flux can quickly decrease whereas the 1500\AA flux will be largely unaffected. Using numerical simulations of galaxies with realistic feedback prescriptions that broadly reproduce the spread in the star-forming main sequence at these epochs \citep{whitaker12}, \citet{dominguez14} show that galaxies with $M^* > 10^9 M_{\odot}$ have a $\sim0.09$ dex (or $\sim 23$\%) spread in the $(f_{1500}/f_{LyC})_{stel}$ ratio. 

In addition, the $(f_{1500}/f_{LyC})_{stel}$ ratio may be significantly lower (by a factor of $\sim2$) when including the effects of the binarity and rotation of massive stars, as the number of WR stars is increased and the main sequence lifetime and effective temperature are increased \citep{eldridge09}. 

The IGM transmission, $\textrm{exp}(-\tau_{IGM,HI})$, is integrated along the entire F336W filter bandpass.  

\begin{equation}
\textrm{exp}(-\tau_{IGM,HI}) = \frac{\int \textrm{exp}(-\tau_{\lambda})\ T_{\lambda}\ d\lambda}{\int T_{\lambda}\ d\lambda}
\label{eqn:igm}
\end{equation}

\noindent where the IGM opacity, $\tau$, and the filter transmission, $T$, are both functions of wavelength. Using a similar technique as in \citet{nestor13} to simulate the IGM lines of sight, the value for the average IGM transmission through the F336W filter curve at $z=3.09$ was estimated to be 0.184 though the value can vary wildly among sight lines (16th and 84th percentile values are 0.011 and 0.365, respectively). The most opaque sight lines can be ruled out if the NB3640 detections are real. The IGM transmission varies slowly with redshift. In Table \ref{tab:props} we list the average transmission value at each redshift and use that value to compute $f_{esc,rel}^{LyC}.$

\begin{center}
\begin{deluxetable*}{lccrrrrrr}
\tablecaption{Galaxy Properties}
\tablehead{\colhead{Galaxy} & \colhead{$z_{spec}$} & exp($-\tau_{IGM}$)\tablenotemark{a} & \colhead{$f_{esc,rel}^{LyC}$\tablenotemark{b}\tablenotemark{c}} & \colhead{F336W\tablenotemark{b}\tablenotemark{d}} & \colhead{F606W\tablenotemark{d}} & \colhead{F814W\tablenotemark{d}} & \colhead{F110W\tablenotemark{d}} & \colhead{F160W\tablenotemark{d}}}
\startdata
C49	&	3.153 & 0.158 & $<0.088$ &  $<0.037$ & $5.74 \pm 0.16$ & $6.37 \pm 0.10$ & $6.69 \pm 0.14$ & $8.13 \pm 0.24$ \\
C49\_NW	&	2.029 & & \nodata &  $0.40 \pm 0.026$ &  $1.73 \pm 0.19$ & $1.28 \pm 0.07$ & $1.89 \pm 0.10$ & $3.28 \pm 0.17$ \\
MD46 &    	3.091 & 0.0183 & $<0.066$ & $<0.060$ & $9.31 \pm 0.25$ & $11.9 \pm 0.16$ & $11.1 \pm 0.18$ & $14.4 \pm 0.33$ \\
MD46\_NW &   \nodata &	& 3.62\tablenotemark{e} & $0.40 \pm 0.022$ &  $0.87 \pm 0.09$ & $1.45 \pm 0.06$ & $1.47 \pm 0.10$ & $1.23 \pm 0.18$  \\
MD32	&	3.102  & 0.178 &  $<0.451$ & $<0.025$  & $0.41 \pm 0.11$ & $0.75 \pm 0.07$ & $0.78 \pm 0.07$ & $1.43 \pm 0.12$ \\ 
MD32\_NE &  2.885? 1.964? & & \nodata & $0.56 \pm 0.038$ & $1.74 \pm 0.16$ & $2.22 \pm 0.10$ & $3.28 \pm 0.11$ & $5.41 \pm 0.20$  \\
aug96M16\_E & 3.095  & & \nodata & \nodata & \nodata &  $2.73 \pm 0.17$ & \nodata & $6.31 \pm 0.45$   \\
aug96M16 & \nodata &  & \nodata & \nodata & \nodata & $3.87 \pm 0.24$  & \nodata & $9.04 \pm 0.49$  \\
aug96M16\_W & 3.291 & & \nodata & \nodata & \nodata &  $3.22 \pm 0.14$ & \nodata & $4.06 \pm 0.32$ \\
D17 & 3.069 & 0.194 &  $<0.081$ & $<0.031$ & $4.10 \pm 0.14$ & $4.78 \pm 0.20$ & $4.61 \pm 0.13$ &  $6.37 \pm 0.22$ 
\enddata
\tablenotetext{a}{Defined in Equation \ref{eqn:igm}. Here we use the average transmission of 500 simulated sightlines.}
\tablenotetext{b}{``$<$" denotes $1\sigma$ limits. }
\tablenotetext{c}{Defined in Equation \ref{eqn:fesc}.}
\tablenotetext{d}{Flux density in units of $10^{-30}$ erg s$^{-1}$ cm$^{-2}$ Hz$^{-1}$.}
\tablenotetext{e}{For this calculation, MD46\_NW was assumed to have the same redshift as MD46, $z=3.091$, and the same value of transmission through the IGM. An $f_{esc,rel} > 1$ means that the measured flux is larger than assumed.  This could be caused by a very recent burst in star formation \citep{dominguez14}.  The number may be artificially high if line of sight transmission is higher than average.} 
\label{tab:props}
\end{deluxetable*}
\end{center}

\section{NIRSPEC Observations}
\label{nirspec:obs}
Near-IR spectra were obtained for four of the targets using the NIRSPEC instrument 
\citep{mclean98} on the Keck II telescope. We observed with NIRSPEC 
on 18 and 19 August 2011. Conditions were photometric, with 
0\secpoint5 seeing in the $K$ band. The exposure times were 
6$\times$900~seconds for C49 and MD32, 8$\times$900~seconds for MD46, 
and 4$\times$900~seconds for aug96M16. All targets were observed with 
a 0\secpoint76 $\times$ 42$''$ long slit.
As explained in Section \ref{sect:results}, many of the targets are either elongated or have several detected clumps (or separate galaxies) within a 1\secpoint2 radius. The position angle of the NIRSPEC slit (listed in Table \ref{tab:nirspec_obs})
was chosen in the direction of elongation or to collect spectra of multiple objects.
Since the primary spectral features of interest were 
[OIII]$\lambda$5007 and H$\beta$ at $z\sim 3.06 - 3.30$, we used the N6 
filter, a broad $\it{H+K}$ filter centered at 1.925$\mu$m with a 
bandwidth of 0.75$\mu$m. The spectral resolution of our NIRSPEC 
observations was $\sim15$ \AA, as determined from sky lines measured 
in the N6 filter. This resolution corresponds to 
$\Delta\upsilon\sim$230 km s$^{-1}$ (R$\sim$1300). Due to the faint 
nature of these objects in the $\it{K}$ band, we acquired each target 
using blind offsets from a bright star in the surrounding field.  We 
returned to the offset star between each integration of the science 
target to recenter and dither along the slit.  In all cases we 
dithered back and forth between two positions near the center of the 
slit.

\subsection{NIRSPEC Data Reduction}
\label{nirspec:redux}
Data reduction was performed using a method similar to that described 
in \citet{liu08}, where the sky background was subtracted from the 
two-dimensional unrectified science images using an optimal method 
\citep[][G. D. Becker 2006, private communication]{kelson03}. The 
sky lines were fit with a low-order polynomial and a b-spline fit was 
used in the dispersion direction. After background subtraction, 
cosmic rays and bad pixels were removed from each exposure.  The 
individual images were then rotated, cut out along the slit, and 
rectified in two dimensions to take out the curvature both in the 
wavelength and spatial directions.  The final rectified 
two-dimensional exposures for each object were then registered and 
combined into one spectrum.

Each of our targets had previously been spectroscopically confirmed as a $z\sim 3$ galaxy based on optical (i.e., rest-frame UV) spectroscopy. Our goal was to determine if the 
specific regions associated with apparent LyC emission 
were also at $z> 3.06$, high enough redshift that the NB3640 and F336W 
filters provided a clean probe of the LyC spectral 
region. Accordingly, in addition to confirming the redshift for the 
main target, we also searched the NIRSPEC slit for additional 
emission-line features offset from the main region of non-ionizing UV 
flux, and perhaps originating from the same regions as the NB3640 or 
F336W flux. 

In the spectra of C49, aug96M16, and MD32, we discovered 
evidence for an additional emission-line spectrum coincident with or 
very near the location of NB3640/F336W emission (referred to, 
respectively, as C49\_NW, aug96M16\_E \& aug96M16\_W, and MD32\_NE), while MD46 
yielded only a single detected spectrum centered on the location of 
the non-ionizing UV continuum.

\begin{figure*}
\epsscale{1.}
\plotone{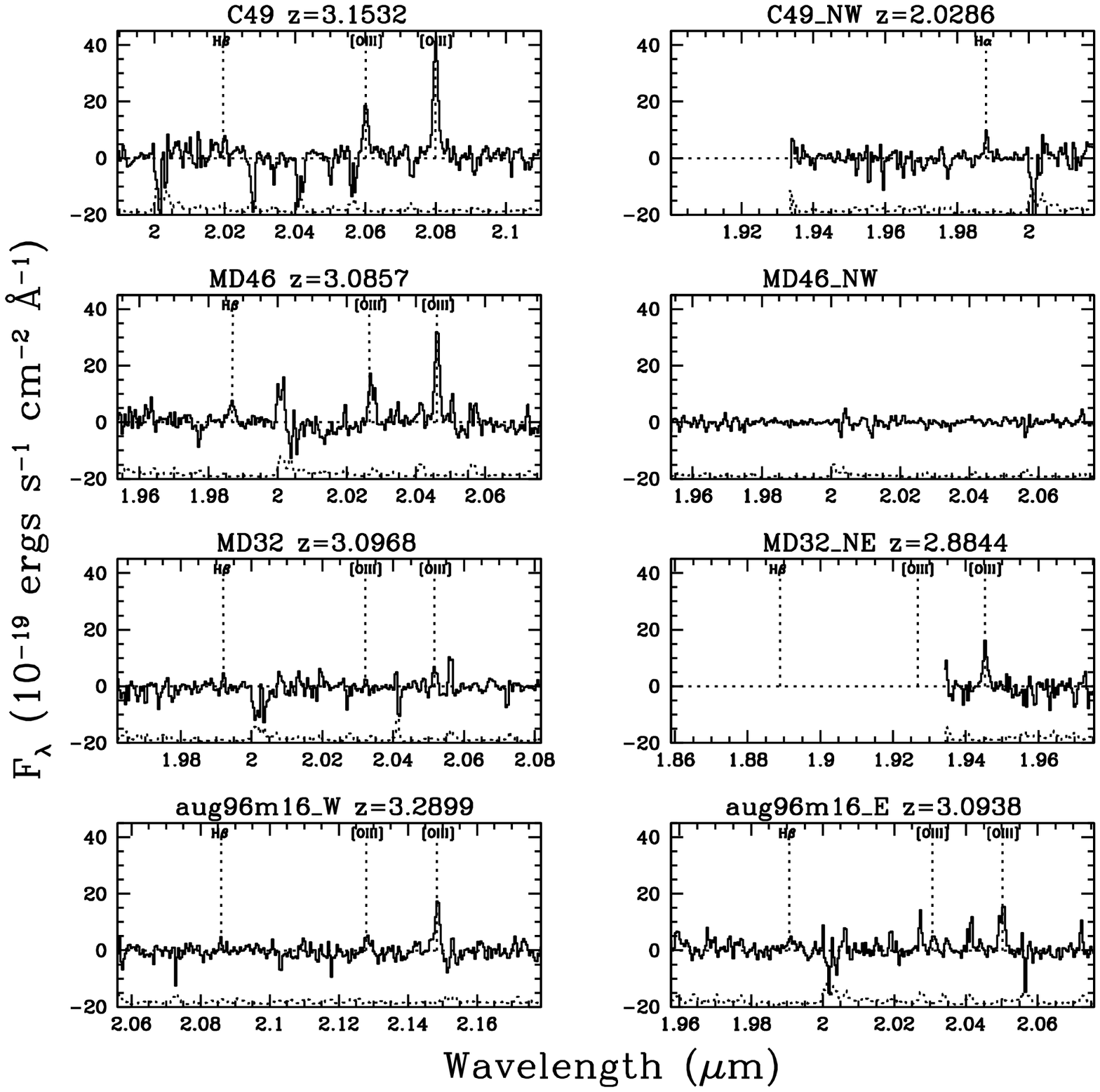}
\caption{The 1D NIRSPEC spectra of all components from which line emission was detected. The $1\sigma$ error spectrum is plotted as a dotted curve, and offset vertically by $20\times 10^{-19}$ ergs s$^{-1}$ cm$^{-2}$ \AA$^{-1}$ from the corresponding science spectrum in each panel. The associated redshifts are given above each spectrum.  The candidate LyC emitter, C49\_NW, is found to be foreground emission from a galaxy at $z=2.0286$. The candidate LyC emitter, MD46\_NW has no detectable emission at the redshift of MD46 ($z\sim3.09$, or any other redshift) so it cannot be verified as a LyC emitter based on these data alone. The candidate LyC emitter, MD32\_NE, is found to have a redshift of $z=2.8844$ if the detected line is [O {\sc iii}]$\lambda5007$.  For aug96m16, the NIRSPEC spectrum show us that our initial redshift was assigned to the incorrect galaxy.  The central galaxy, which exhibits emission in the Keck NB3640 filter is not at z=3.2899.  No redshift has been definitively identified for the candidate LyC emitter.}
\label{fig:nirspec1d}
\end{figure*}

One-dimensional spectra were extracted from the two-dimensional 
reduced image at the location of any detected emission lines. 
The corresponding 1$\sigma$ error spectra were also extracted.  The 
average aperture size along the slit was 1.$^{\prime\prime}$3, with a 
range of 1.$^{\prime\prime}$1 to 2.$^{\prime\prime}$0.  Each 
one-dimensional spectrum was then flux-calibrated using A-type stars 
according to the method described in \citet{shapley05} and 
\citet{erb03} and placed in a vacuum, heliocentric frame. The final flux-calibrated, one-dimensional spectra are plotted in Figure \ref{fig:nirspec1d}.

\subsection{NIRSPEC Measurements}
\label{nirspec:meas}

In the rest-frame optical, the 
strongest features are [OIII]$\lambda 5007$ and H$\beta$. 
We determined 
the [OIII]$\lambda$5007 centroid, flux and FWHM, by fitting a 
Gaussian profile to each emission line using the $\tt{IRAF}$ task, 
$\tt{splot}$. The nebular velocity dispersion was calculated as 
$\sigma_v({\mbox{O[III]}})$ $=$ (FWHM/2.355) 
$\times(c/\lambda_{obs})$, where $\lambda_{obs}$ is the observed 
wavelength of [OIII]$\lambda$5007.  The FWHM was determined from the 
subtraction of the instrumental FWHM (estimated from the widths of 
night sky lines) from the observed FWHM in quadrature. Uncertainties 
in the velocity dispersion were determined using the same method 
described in \citet{erb06b}.  Where detected, we also measured the properties of H$\beta$ emission.

A Monte Carlo approach was used to measure the uncertainties in the 
emission-line centroid, flux, and FWHM.  For each object 500 
simulated spectra were created by perturbing the flux at each 
wavelength of the true spectrum by a Gaussian random number with the 
standard deviation set by the level of the the 1$\sigma$ error 
spectrum.  Line measurements were obtained from the fake spectra in 
the same manner as the actual data. The standard deviation of the 
distribution of measurements from the artificial spectra was adopted 
as the error on each centroid, flux, and FWHM value. Upper limits on the H$\beta$ line fluxes were obtained for non-detections by assuming the same FWHM as the [OIII]$\lambda$5007 emission line and performing the same Monte Carlo measurements as described above. Based on [OIII]$\lambda$5007 emission line centroid measurements, we obtained the systemic redshifts for each object.  In one case (MD32\_NE) the single detected emission line was assumed to be [O {\sc iii}]$\lambda$5007. If this line identification is correct, [O{\sc iii}]$\lambda$4959 and H$\beta$ would not be covered by wavelength range of that spectrum. All NIRSPEC measurements are listed in Table~\ref{tab:nirspec_obs}.

\begin{deluxetable*}{lrccccccc}
\tablewidth{0pt} \tabletypesize{\footnotesize}
\tablecaption{NIRSPEC Measurements
\label{tab:nirspec_obs}}
\tablehead{
\colhead{Object} &
\colhead{Slit PA\tablenotemark{a}} &
\colhead{$z_{em}$\tablenotemark{b}} &
\colhead{H$\alpha$ Flux\tablenotemark{c}} & 
\colhead{H$\alpha$ FWHM\tablenotemark{d}} &
\colhead{[OIII] Flux\tablenotemark{c}} &
\colhead{[OIII] FWHM\tablenotemark{d}} &
\colhead{$\sigma_v({\mbox{[OIII]})}$\tablenotemark{e}} &
\colhead{H$\beta$ Flux\tablenotemark{c}}
}
\startdata
C49$\tablenotemark{f}$         & $-38$ & 3.1532 & \nodata & \nodata & 7.5 $\pm$ 0.25 & 17 $\pm$ 0.7 & 54.7$^{+8}_{-9}$ & $<$1.4 \\
C49\_NW                        & $-38$ & 2.0286 & 1.4 $\pm$ 0.16 & 14 $\pm$ 2.6 & \nodata & \nodata & \nodata & \nodata \\
MD46                           & $-34$ & 3.0857 & \nodata & \nodata & 5.6 $\pm$ 0.15 & 16 $\pm$ 0.5 & 27.1$^{+10}_{-15}$ & 1.4 $\pm$ 0.15 \\
MD46\_NW\tablenotemark{f}  & $-34$ & \nodata & \nodata & \nodata & $<$0.40 & \nodata & \nodata & \nodata \\
MD32$\tablenotemark{f}$    & 37& 3.0968 & \nodata & \nodata & 1.2 $\pm$ 0.14 & 14 $\pm$ 1.4 & \nodata & $<$0.36 \\
MD32\_NE                   & 37 & 2.8844\tablenotemark{g} & \nodata & \nodata &  2.3 $\pm$ 0.19 & 15 $\pm$ 1.7 & \nodata & \nodata \\
aug96m16\_W\tablenotemark{f} & 90 & 3.2899 & \nodata & \nodata & 3.0 $\pm$ 0.24 & 15 $\pm$ 1.8 & $<$50.48 &  $<$0.82 \\
aug96m16\_E                    & 90 & 3.0938 & \nodata & \nodata & 2.9 $\pm$ 0.34 &  18 $\pm$ 2.1  & 60.2$^{+22}_{-31}$ & 0.8 $\pm$ 0.25 \\
aug96m16\tablenotemark{f} & 90 & \nodata & \nodata & \nodata & $<$1.2 & \nodata & \nodata & \nodata 
\enddata
\tablenotetext{a}{Slit position angle in degrees East of North.}
\tablenotetext{b}{Rest-frame optical nebular emission-line redshift measured from [OIII]$\lambda$5007 or H$\alpha$.}
\tablenotetext{c}{Emission-line flux and error in units of $10^{-17}$ ergs s$^{-1}$
cm$^{-2}$.}
\tablenotetext{d}{Observed FWHM and error in units of \AA.  The FWHM is a raw value,
uncorrected for instrumental broadening.}
\tablenotetext{e}{Nebular velocity dispersion measured in units of km s$^{-1}$.  }
\tablenotetext{f}{Upper limits for [OIII] and H$\beta$ fluxes are quoted at the 3$\sigma$ level.
For aug96M16 and MD46\_NW, limits were calculated by assuming the same
redshift, respectively as aug96M16\_E ($z=3.0938$) and MD46 ($z=3.0857$).}
\tablenotetext{g}{Redshift estimated assuming that the single emission line detected
corresponds to redshifted [OIII]$\lambda 5007$. If the emission line is instead redshifted
H$\alpha$, the redshift MD32\_NE is $z=1.9635$.}
\end{deluxetable*}

\section{Results}
\label{sect:results}

Because there are only five galaxies in our sample and each has different properties, we have chosen to discuss the objects individually. We first discuss the {\it HST} imaging, as the resolved images informed the strategy for the near-IR spectroscopy (slit orientation and required seeing for each galaxy).  In all cases one must keep in mind that the Keck narrow-band filter (NB3640) used to identify the LyC emission has a width corresponding to $\sim 25$ physical Mpc, significantly smaller than the mean free path of a photon at the Lyman limit at $z\sim3.09$ \citep[$\sim100$ Mpc,][]{prochaska09,omeara12}.  The {\it HST} F336W filter, which we have used to detect the LyC of these galaxies, has a similar red cutoff (near the Lyman limit) but is a factor of $\sim5.5$ times wider, with the additional transmission blueward of the NB3640 filter (Figure \ref{fig:spec_filt}).  At the bluer wavelengths probed with the F336W filter, but not probed by the NB3640 filter, the average line-of-sight through the IGM toward a $z=3.09$ galaxy will have significantly less transmission. Thus, the typical galaxy at $z=3.09$ should appear significantly fainter (1.0 magnitudes) in the F336W filter than in the Keck NB3640 filter (see Section \ref{sect:discussion} for more details).  

\begin{figure*}
\epsscale{1.1}
\plotone{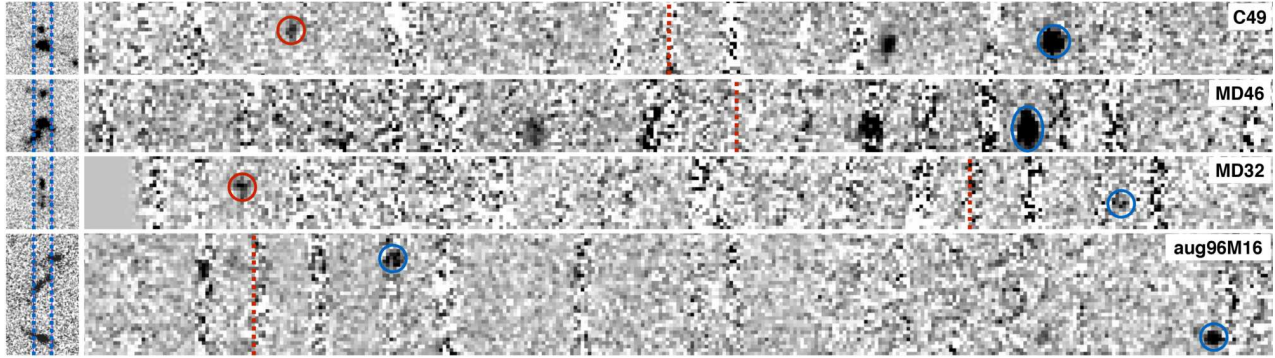}
\caption{The 2D NIRPSEC spectra (right) and the $HST$ F814W stamps of the targets (left) rotated so that the NIRSPEC slit is oriented up-down, to match the 2D spectra. The y-direction spans 3$''$ for the top three galaxies, 5$''$ for the bottom galaxy and the spectra and images have the same spatial scale. The red dashed vertical line denotes the wavelength above which the [O{\sc iii}]$\lambda$5007 line must lie in order for the NB3640 and F336W filters to measure purely LyC, without contamination from flux at wavelengths above the Lyman limit. Both C49 and MD32 show emission lines at two redshifts with the lower redshift lines coming from the higher (in the y direction) of the two components (which are the components detected in NB3640 and F336W).  In MD46 there is no detection of emission lines from the faint companion which is detected in NB3640 and F336W.  Even if the galaxy were at the same redshift, we would be able to resolve [O{\sc iii}]$\lambda$5007 emission from it because the separation of the two components (1\secpoint2) is much larger than the seeing (0\secpoint5).  Finally, we detect emission lines for two of the galaxies in the region of aug96M16 but, given the vertical separation of the lines, they are associated with the upper and lower galaxies. We have not identified a redshift for the central galaxy, which is the one of interest as it is detected in the NB3640 and F336W filters.}
\label{fig:nirspec2d}
\end{figure*}

\subsection{C49}

The galaxy C49 was first identified as a LyC emitter in deep rest-frame ultraviolet spectroscopy from Keck \citep{shapley06}, and again with narrow-band imaging just below the Lyman limit \citep{iwata09, nestor11}.  At ground-based resolution (0\secpoint8), there was already a suggestion in Figure 3 of \citet{nestor11} that the LyC flux was only emitted from the NW corner of the non-ionizing UV-emitting area of C49.

The {\it HST} imaging resolves C49 into two sources separated by 0\secpoint65 (5.0 proper kpc at $z=3.153$) in all filters. The fact that the two clumps are distinct at rest-frame optical wavelengths suggests that these are two separate galaxies rather than two UV-bright star-forming regions of the same galaxy.  Because the galaxy to the SE is brighter and has a previously known redshift ($z=3.153$), we are referring to it with the original C49 label and the nearby galaxy as C49\_NW.  \citet{shapley06} had previously obtained a deep optical spectrum (rest-frame UV) of this system, but the two sources were unresolved and most of the light from both objects fell within the 1\secpoint2 wide slit.  Because C49 is five times brighter than C49\_NW in the optical, it is the galaxy with the previously reported redshift.  

The F336W (LyC) image shows strong emission from C49\_NW, with no detectable emission from C49. We derive a $1\sigma$ limit of $f_{esc,rel} < 0.088$ ($1\sigma$) for C49. If C49\_NW is at the same redshift as C49, or at any other redshift above $z=3.06$, then the F336W flux would be a direct detection of LyC emission.

The NIRSPEC slit was oriented to lie along both objects ($PA=-38^{\circ}$).  The seeing was sufficient (0\secpoint5) to marginally resolve [O {\sc iii}] lines from both components if they were both at the same redshift.  The 2D spectrum is shown in Figure \ref{fig:nirspec2d}.  The strong [O{\sc iii}] $\lambda\lambda4959,5007$ emission lines are clearly detected at the redshift of C49.  There is no indication of fainter [O{\sc iii}] emission at the same redshift from the companion.  Instead, an emission line is detected at shorter wavelength ($\lambda = 1.988$ $\mu$m) $\sim0\secpoint5$ from C49 in the direction of C49\_NW.  The two most likely identification of this line are [O{\sc iii}]$\lambda$5007 or H$\alpha$, indicating redshifts of $z=2.9698$ or $z=2.0286$, respectively. 

\begin{center}
\begin{figure*}

\epsscale{1.2}
\plotone{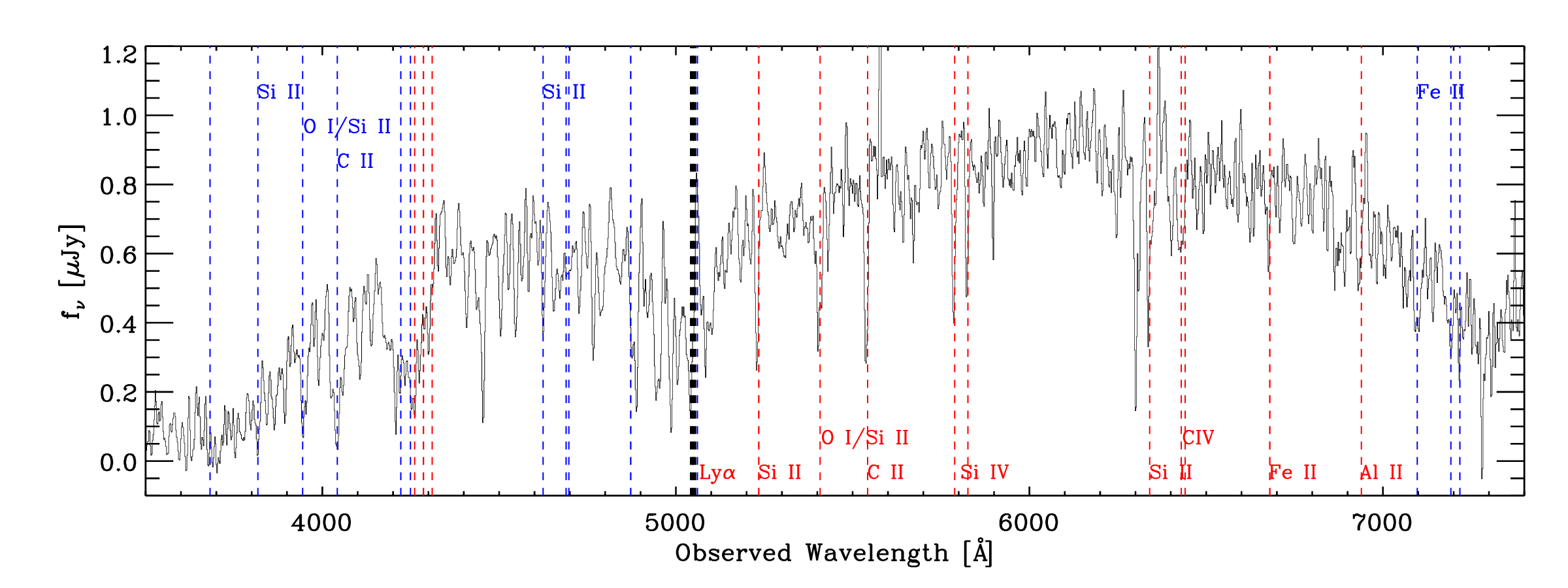}
\caption{LRIS spectrum of C49 (and C49\_NW in the same slit) from \citet{shapley06}. Red dashed lines denote the wavelengths of strong UV absorption lines at the redshift of C49 ($z=3.1532$, strongly detected lines labelled in red at the bottom of the plot).  The blue dashed lines denote those same lines at $z=2.0286$ (strongly detected lines labelled in blue at the top of the plot). Within the Ly$\alpha$ forest at $\lambda < 5049$ \AA (denoted with a thick black dashed line), we still see strong absorption coinciding with Si {\sc ii} $\lambda 1260$, O {\sc i} / Si {\sc ii} $\lambda\lambda 1302, 1304$, C {\sc ii} $\lambda 1334$, Si {\sc ii} $\lambda 1526$, and Fe {\sc ii} $\lambda\lambda\lambda 2343, 2374, 2382$. Therefore, we conclude that the emission line detected in the NIRSPEC spectrum for C49\_NW is H$\alpha$, and C49\_NW has z=2.0286. The redshift of C49\_NW would have been difficult to determine with this LRIS spectrum alone because all of the strong absorption features are in the Ly$\alpha$ forest or at the red (low S/N) edge of the spectrum.}
\label{fig:c49_lris}

\end{figure*}
\end{center}

Given that C49\_NW is likely at one of these two redshifts, we re-examined the deep LRIS spectrum of C49 \citep{shapley06} to look for absorption features from a foreground source. As mentioned above, the impact parameter is $\sim 5$ kpc, and strong absorption from both the interstellar and circumgalactic medium of C49\_NW would be expected \citep{steidel10}.  In Figure \ref{fig:c49_lris}, we plot the deep Keck/LRIS spectrum. There are several strong absorption features associated with $z=2.0286$ in the Ly$\alpha$ forest (Si {\sc ii} $\lambda 1260$, O {\sc i} / Si {\sc ii} $\lambda\lambda 1302, 1304$, C {\sc ii} $\lambda 1334$, Si {\sc ii} $\lambda 1526$) and at the red end of the spectrum (Fe {\sc ii} $\lambda\lambda\lambda 2343, 2374, 2382$). 
Therefore, we conclude that the emission line in the NIRSPEC spectrum is H$\alpha$ and the flux detected at $\lambda \lesssim 3780$ \AA\ in both \citet{shapley06} and \citet{nestor11} is not LyC emission. The foreground contamination was not originally recognized with the LRIS spectrum alone because all of the strong UV absorption features are in the Ly$\alpha$ forest of C49 (except for the faint Fe {\sc ii} lines at the red end of the spectrum where the sensitivity is poor). The additional detection of an emission line in the rest-frame optical allowed us to identify the absorption features as associated with C49\_NW and not with foreground Ly$\alpha$ lines.

\subsection{MD46} 

MD46 is a bright LBG at the redshift of the overdensity ($z=3.09$) and was identified as a candidate LyC emitter by \citet{iwata09} and \citet{nestor11}. \citet{iwata09} suggested that the LyC was offset from the brightest region of MD46. The {\it HST} images of the rest-frame non-ionizing UV and optical continuum show several clumps.  As seen in Figure \ref{fig:stamps}, most of the emission is coming from two bright clumps and two fainter clumps all within a 0\secpoint7 diameter region, which we are collectively calling MD46. It is this bright cluster of objects that has the known spectroscopic redshift. There is an additional object 1\secpoint2 (9.2 kpc) to the NW that we will refer to as MD46\_NW.  MD46 is not detected in the {\it HST} F336W (LyC) image but MD46\_NW is strongly detected. We derive a $1\sigma$ limit of $f_{esc,rel} < 0.066$ ($1\sigma$) for MD46. If MD46\_NW has the same redshift as the brighter counterpart to the SE (or any redshift above $z=3.06$), then the detected F336W flux is LyC emission.  

As seen in Figure \ref{fig:stamps} the Ly$\alpha$ emission is clearly emitted from the NW side of MD46, in the direction of MD46\_NW, possibly indicating that MD46\_NW is at the same redshift. \citet{nestor13} use the Ly$\alpha$ image and a newly-obtained deep rest-frame UV spectrum covering Ly$\alpha$ showing that the Ly$\alpha$ emission extends from MD46 in the N and W directions and coincides with MD46\_NW. They therefore conclude that MD46\_NW is likely at the same redshift as the brighter MD46. Additional evidence for this interpretation is that there are no unexplained strong absorption lines in the high $S/N$ rest-UV spectrum of MD46, which would arise from passage through the halo (within $\sim 10$ kpc) of MD46\_NW if it were in the foreground. However, neither of these arguments (the extended Ly$\alpha$ emission or the lack of absorption in the MD46 spectrum) definitively proves that MD46\_NW is at the same redshift as MD46.  First, as seen in C49, if the foreground object is at the right redshift ($z\sim2$), most strong absorption features will either be in the Ly$\alpha$ forest or fall off the red end of the spectrum.  Also, because Ly$\alpha$ is a resonance transition, Ly$\alpha$ emission is commonly seen extended beyond the stellar extent (and often asymmetric).  Thus, a definitive confirmation must come from a non-resonance, emission line such as [O {\sc iii}]$\lambda$5007.  

The NIRSPEC slit was oriented to obtain spectra of both objects, MD46 and MD46\_NW ($PA = -34^{\circ}$). The separation between the two components ($1\secpoint2$), is large enough that emission lines at the same wavelength from the two targets would easily be resolved if at the same redshift.  [O {\sc iii}]$\lambda\lambda$4959,5007 and H$\beta$ from MD46 are detected at high significance ($S/N=37$ in the case of [O {\sc iii}]$\lambda$5007).  However, no emission lines are seen at the location of MD46\_NW, with a $3\sigma$ upper limit of $4.0\times10^{-18}$ erg s$^{-1}$ cm$^{-2}$. 

We can estimate the expected [O {\sc iii}]$\lambda$5007 emission line flux for MD46\_NW to determine whether our observations are sensitive enough to detect the line {\it if} it lies at the same redshift as MD46.  We make the assumption that the ratio of the [O {\sc iii}]$\lambda$5007 lines of the two galaxies is the same as the ratio of the rest-frame UV fluxes (as seen in the F814W filter). This is only true if the [O {\sc iii}] emission traces the star formation rate and if there is similar excitation and dust extinction in both galaxies.  The ratio of F814W fluxes is 8.21, giving a predicted [O {\sc iii}]$\lambda$5007 flux for MD\_NW of $6.8\times10^{-18}$ ergs s$^{-1}$ cm$^{-2}$, which would be a $5.1\sigma$ detection. Thus, the lack of detection suggests that MD46\_NW might be at a different redshift such that its strong lines lie outside of the K-band (or on a sky line).  Of course, this argument assumes that the [O {\sc iii}] emission is proportional to the rest-UV continuum emission, but the $L_{[OIII]}/L_{UV}$ ratio is a function of star formation history, metallicity, and dust extinction.  One might expect that the lower mass galaxy (MD46\_NW) has a lower metallicity and, thus, a higher [O {\sc iii}] to Balmer line ratio \citep[eg. ][]{erb06a,dominguez13}, indicating that its [O {\sc iii}] to star formation ratio should be high.  However, MD46\_NW is redder in F606W-F814W color, suggesting that it may be dustier.  Unfortunately, given the dispersion in UV-to-[O {\sc iii}] flux ratios seen in star-forming galaxies, it is impossible to definitively determine with these data alone whether or not MD46\_NW is at the same redshift as MD46 and, therefore, whether or not it is a LyC emitter.

\subsection{MD32}

MD32 is the faintest target in our sample and is an elongated galaxy with an extent of $\sim 1\secpoint2$, which appears to be edge on.  The F814W image shows two bright clumps to the NE and SW and a fainter, smaller clump between them ($\sim$ 0\secpoint2 of space between the two regions).  The NE region is brighter and strongly detected in the F336W (LyC), as is the faint clump near the center. The SW clump is undetected in the F336W image.  

The NIRSPEC slit was placed in the direction of elongation of MD32 ($PA=37^{\circ}$) to search for emission from the two components or to identify rotation in what could be an edge-on disk. Faint line emission at the expected wavelength of [OIII]$\lambda$5007 at $z=3.0968$ is detected, confirming the redshift determined via identification of Ly$\alpha$ emission detected in the LRIS spectrum \citep{nestor13}.  We also detect a stronger emission line at shorter wavelengths ($\lambda = 1.945$ $\mu$m).  This line is offset nearly $1''$ along the slit in the NE direction, corresponding to the region of MD32 detected in the F336W image. Therefore, we conclude that the F336W-detected portion is at a lower redshift ($z=2.8844$ or $z=1.9635$ if the line is [O {\sc iii}]$\lambda$5007 or H$\alpha$, respectively), and is therefore not escaping LyC.  Because MD32 has long been identified as being at $z=3.09$, we refer to the SW clump  as MD32 and the NE clump as MD32\_NE. We derive a $1\sigma$ limit of $f_{esc,rel} < 0.451$ ($1\sigma$) for MD32.

The importance of high-resolution imaging combined with rest-frame optical spectroscopy is demonstrated well here. The {\it HST} F336W image allowed us to determine definitively from which clump the NB3640 emission was emitted, and the rest-frame optical spectrum determined that the NB3640-detected clump is at lower redshift.  

\subsection{aug96M16}

The object aug96M16 is perhaps the most complex of our targets as there are three galaxies lined up in the E-W direction. aug96M16 does not fall within the F336W footprint, but we have reduced existing $HST$ F814W and F160W imaging of this system. The center and eastern galaxies are confused at ground-based resolution, but the $HST$ images reveal that there are three distinct galaxies.  We will refer to these objects, from East to West, as aug96M16\_E, aug96M16, and aug96M16\_W. 

The centroid of the NB3640 coincides with the middle of the three galaxies, aug96M16. This galaxy was erroneously assigned a redshift of $z=3.29$ in \citet{nestor11} but \citet{nestor13} corrected this and identified the westernmost galaxy, aug96M16\_W as having this spectroscopic redshift.  Thus, aug96M16 should not have been included as a candidate LyC emitter because its redshift was unknown. Our NIRSPEC spectra were obtained before this error was realized. Nonetheless, if the central galaxy is at $z>3.06$, the detected NB3640 emission would be LyC emission.

The NIRSPEC slit was aligned E-W (PA$=90^{\circ}$) so that all three objects fell on the slit. Two sets of [O {\sc iii}]$\lambda$5007,$\lambda$4959 emission lines are detected in the 2D spectrum, corresponding to redshifts of 3.0938 and 3.2899.   The latter redshift agrees with the redshift reported in \citet{nestor11}.  If the galaxy emitting flux in the NB3640 were at either of these redshifts, then the NB3640 would be LyC emission.  Because we don't know the absolute position of the slit on the sky, we use the distance between the objects to discern which lines correspond to which galaxies. The physical separation along the slit is 3\secpoint24.  This corresponds to the separation between the easternmost and westernmost galaxy. Thus, the westernmost galaxy has a spectroscopic redshift of 3.2899 \citep[consistent with ][]{nestor13} and the easternmost galaxy has a spectroscopic redshift of z=3.0938.  We did not detect any strong emission lines from the central galaxy that is detected in the NB3640 filter. Thus, the central galaxy can not be confirmed as a definitive LyC emitter.  

\subsection{D17}

D17 is a bright compact galaxy located in the protocluster at $z=3.09$, and was identified as having diffuse LyC emission offset $\sim1''$ to the NE by \citet{nestor11}. There is not a galaxy detected at the location of the emission at longer wavelengths.  However, it is physically possible to have nebular LyC emission from a cloud that is photo-ionized by LyC escaping from a nearby galaxy \citep{inoue10}. However, such models require extreme stellar populations \citep[$<1/50$ $Z_{\odot}$, age $< 1$ Myr,][]{inoue11}, and one of the best cases for such nebular LyC emission \citep[object `a' in][]{inoue11} was found to actually be due to a foreground galaxy contaminating the spectral energy distribution \citep[LAE 003 in][]{nestor13}. 

Because D17 lies in our F336W image footprint, we investigated whether we also detect this diffuse emission in our image. The reported NB3640 emission near D17 was just above the detection limit, with an AB magnitude of 27.0. Assuming an average line-of-sight through the IGM and a flat (in $f_{\nu}$) LyC spectrum, we would expect the F336W magnitude to be $28.0$ (AB). At 1 arcsec$^2$, that flux level corresponds to $4.3\sigma$. There is no significant detection in the F336W image. However, because the IGM transmission can vary significantly with wavelength (it can be transparent at 3640 \AA\ and opaque at shorter wavelengths), the F336W non-detection does not invalidate the detection in NB3640. Therefore, our image can not confirm the existence of offset nebular continuum NE of D17. 

We did however use our F336W image to estimate the LyC escape fraction ``down-the-barrel" toward D17. Using an aperture defined by the D17 near-IR image, we derive a $1\sigma$ limit of $f_{esc,rel} < 0.081$.

Because there are not other galaxies around D17 and the one galaxy has a known redshift, a NIRSPEC spectrum was not obtained for this system.  

\section{Discussion}
\label{sect:discussion} 

There have been several surveys for LyC-emitters at this redshift, some detecting few or no LyC-emitters \citep{boutsia11,vanzella12}, some claiming dozens of candidates \citep{iwata09,nestor11,nestor13,mostardi13}.  It can be confusing to synthesize these results into a unifying picture.  As we have seen in this paper, it is critical to carefully consider contamination and its effects on each survey when trying to compare results.  

The rate of contamination from foreground galaxies is quite low at these wavelengths.  As an example, the odds of a foreground galaxy with $U_{AB}<27.5$ lying within 1\secpoint0 of a target galaxy at $z\sim3$ is roughly 8.5\% given the completeness-corrected surface density tabulated in \citet{vanzella10} and taken from \citet{nonino09}.  But detection of LyC emitters is also rare.  Obviously, if the detection rate is near the contamination rate, then most or all of the candidate LyC emitters will be spurious. It is therefore critical that any survey for LyC-emitting galaxies accurately calculates the contamination rate and has a significantly higher detection rate if it is to claim a bona-fide LyC detection. We list here several factors that can affect the contamination rates and detection rates in these surveys. First, the factors affecting the contamination rate, many of which have been outlined before by other authors \citep{siana07,vanzella10}.  

\begin{itemize}
\item Depth: With increasing depth, fainter and more numerous foreground galaxies \citep[e.g.][]{alavi14} can be detected.
\item Spatial resolution: At lower resolution, foreground galaxies at larger angular separation from the target galaxy can contaminate the photometry.
\item Redshift: At higher redshift, because the line-of-sight to the target galaxy is much larger, the volume of the contaminating cylinder is larger and the number of contaminants will therefore be larger. This is part of the reason that contamination concerns are much smaller for the $z\sim1$ samples of \citet{siana07,siana10} and \citet{bridge10}.
\end{itemize} 

Generally, one can observe the surface density of objects in the field at the wavelengths of the LyC observations, and determine the rate at which an interloper would lie in the the seeing disk. The example above, because it is very deep and uses a fairly large, $1''$ radius aperture, results in an 8.5\% contamination rate.  Of course, one also has to consider the flux distribution of expected interlopers as well.  

Below are the factors affecting the detection rate of galaxies.

\begin{itemize}
\item Depth: Deeper data will be sensitive to lower escape fractions, increasing the detection rate. Deeper data are also sensitive to escaping LyC from less luminous galaxies. 

\item Redshift and IGM opacity:  This factor is often overlooked.  When looking at rest-frame wavelengths near the Lyman limit (e.g. $\lambda_{rest}=850-910$ \AA) at $z \sim 3$, the IGM is relatively transparent.  At shorter rest-frame wavelengths, the IGM transmission is significantly lower. The reason for this difference is that the mean-free-path of a LyC photon at $z\sim3$ is $l_{mfp} \sim 100$ Mpc \citep{prochaska09}. As a photon travels that distance, it redshifts by about 10\%.  Therefore, a photon with $\lambda_{rest}=829Ê$ \AA\ at $z\sim3$ has to travel $1l_{mfp}$ to redshift beyond the Lyman limit so that it is no longer subject to bound-free absorption from hydrogen. A photon of $\lambda_{rest}=760$ \AA\ has to travel $2l_{mfp}$ (Note, we are ignoring the much smaller effects of decreasing cross section with decreasing wavelength and the changing mean-free-path with redshift to illustrate the simple point.) In summary, surveys that probe the LyC of a large number of galaxies at an optimal redshift for the filter \citep[e.g.][at $z\sim3.09$]{nestor11} will likely have a higher detection rate than a survey of field galaxies at a variety of redshifts.  Of course, if the observations are done spectroscopically, then one can analyze many galaxies at a variety of redshifts with good sensitivity near the Lyman limit.   
\item Filter width: Because the mean-free-path of a LyC photon in the IGM at $z\sim3$ corresponds to observed 330 \AA\ (see above), the filter width can affect the level of signal detected.  Filters significantly smaller than the mean-free-path (and near the Lyman limit) will probe regions of the LyC with high transmission. Broadband filters however, will probe wavelengths with little or no transmission at the blue end of the filter, as the typical filter (e.g. the Johnson U filter) is twice the width of the mean free path. In fact, a wide filter will often be adding noise at the short wavelength end, but no signal.  Therefore, if two images reach the same depth in AB magnitudes but one uses a narrow band (near the Lyman limit) and the other a broad band, the narrow band will be more sensitive to escaping LyC.
\end{itemize}

All of these factors must be carefully considered when comparing candidate detection rates and contamination rates. If the detection rate is not significantly higher than the expected contamination rate, then no detection should be claimed.  

\citet{nestor11} carefully calculate the contamination rate using the surface density of objects {\it at the same wavelength} as the LyC filter, and using a contamination radius (1.2$''$) equal to the radius used to match the target catalog to the LyC filter catalog.  When doing so, they determine a contamination rate that would account for about half of the detected brighter Lyman break-selected galaxies, and $\sim16$\% of the fainter, Ly$\alpha$-selected galaxies (LAEs). Our results in this paper are consistent with the high contamination rate for the brighter LBGs.

For comparison, \citet{iwata09} covered a much wider area, but 0.6 magnitudes shallower in depth.  Because of this, they were sensitive only to the escaping LyC in more luminous galaxies.  They detect a very low fraction of their candidates, 8.6\% (17 of 198), and estimate that only $\sim2.37$\% ($\sim 4.7$ detections) could be explained by foreground contamination.  However, the contamination rate estimate was not precise.  They used published number counts at 3000 \AA, not the wavelength of their observations ($\sim 3590$ \AA). As number counts rise quickly from the near-UV to optical, the counts turn out to be significantly different.  Also, the counts they use are incomplete, and are therefore too low.  If instead we use the completeness-corrected cumulative number counts at $\sim 3750$ \AA\ from \citet{nonino09, vanzella10}, then we get a much larger foreground source density ($\sim 3.5\times10^5$ deg$^{-2}$ instead of $1.0\times10^5$ deg$^{-2}$).  This estimate increases the expected contamination by a factor of 3.5 and the expected number of contaminants is $\sim 16.6$, consistent with the 17 detected sources. It is true that, because the number counts stated above are from slightly redder wavelengths, the foreground source counts are slightly higher than at $\sim 3590$ \AA.  However, the difference is likely small because of the small wavelength change, and even if the contamination rate were increased by only a factor of 2.5 (to 5.9\%, or 11.7 contaminants), instead of 3.5, from their original estimate, the significance of 17 detections would only be a 1.4 $\sigma$ deviation from the expectation value. Of course, there could be real LyC sources in the sample, but it cannot be claimed with high confidence that the sample is anything but foreground sources.  

\citet{boutsia11} were careful to only analyze galaxies at redshifts where the Lyman limit was close to the red edge of transmission in the filter ($3.27<z<3.35$, when the transmission edge corresponds to $z\sim3.27$).  However, their sample was limited to relatively luminous galaxies (10 of 11 galaxies have $L>L^*$), where the average escape fraction appears to be quite low. In addition, the sample is small, and it is very possible, given the rare LyC detections of luminous galaxies, that had they sampled significantly larger samples, they would have detected some galaxies.  Nestor et al. only detected 20\% of luminous LBGs with deeper data ($1\sigma=29.1$ AB in the narrow-band filter, which is more sensitive to LyC vs. 29.3 AB in the broadband filter), so a null result for \citet{boutsia11} is not unexpected.  

\citet{vanzella10b} had large samples and were careful about contamination.  However, the sample is very bright (so a low percentage of detections is already expected) and many of the galaxies are at much higher redshift than the most sensitive redshifts.  Specifically, less than half of the non-AGN ($58$ of 128) are at $z<3.6$, such that the longest wavelength of transmission in the filter ($\sim 3900$ \AA) is sensitive to the galaxies' LyC emission within one mean-free-path of the \citep[$\l_{mfp}(z=3.6)\sim50$ Mpc, ][]{prochaska09}. Thus, most of these galaxies would not be detectable with significant escape fractions.  In addition, the escape fraction limits for the galaxies at $z\sim3.5-3.6$ are significantly less sensitive than those at $z\sim3.4$, the lowest redshift probed by their filter.  Therefore, it is perhaps not surprising that \citet{vanzella10b} do not have a significant number of detections (after using high resolution $Hubble$ resolution to minimize foreground contamination).  

In summary, all of these surveys at $z\sim3$ are consistent with the interpretation that high escape fractions in bright, $L \gtrsim 0.5 L^*$, galaxies are rare.  Our results in this paper are consistent with this scenario.  In addition, \citet{nestor13} suggest it is more common to have high escape fractions from faint Ly$\alpha$ emitters. In a subsequent paper, we will investigate escape fractions and foreground contamination in the fainter sample.



\section{Conclusions}
\label{conclusions}

In total, there were six bright, $z\sim3.1$, LBGs with candidate LyC emission identified via detection in the Keck/LRIS NB3640 filter \citep{nestor11}.  Here we have presented follow-up $HST$ and Keck/NIRSPEC observations of five of the six candidates.  In each case, the candidate LyC emission is associated with one of multiple clumps or, in the case of D17, offset from the primary galaxy but not obviously associated with another galaxy.  Therefore, spectroscopic follow-up was required to confirm the redshifts of the clumps that were emitting the candidate LyC emission.  

One of the five target galaxies (aug96m16) was found to have been misidentified as a galaxy at $z>3.06$ from the Keck/LRIS optical spectrum. The actual redshift of the galaxy detected in the NB3640 filter is unknown so it should not have been included in the original sample as a LyC-emitting candidate. Therefore, there are four galaxies with $z>3.06$ with candidate LyC emission. The Keck/NIRSPEC spectroscopy revealed that two of the remaining four candidates were found to be contaminated by galaxies at lower redshifts (C49\_NW and MD32\_NE). A redshift for the component to the NW of MD46, MD46\_NW, could not be confirmed.  \citet{nestor13} suggest that extended Ly$\alpha$ emission in the direction of MD46\_NW indicates that the galaxy is likely at the same redshift as the brighter counterpart, MD46. However, the NIRSPEC limit on [O {\sc iii}]$\lambda$5007 is quite low, suggesting that it must have a significantly lower $L_{[O {\sc III}]}/L_{UV}$ ratio than MD46 if MD46\_NW is indeed at the same redshift as MD46. This remains the most promising LyC-emitting candidate of the bright galaxies in this paper.  Finally, the purported LyC emission from D17 is offset by $\sim1''$.  There is not a counterpart galaxy at the location of the emission down to F814W $\sim27.3$ (AB). The low $S/N$ emission seen from the ground is not seen in the F336W image, however, varying IGM opacities may account for this difference.  

In summary, we find that the contamination rate of this bright, $L>0.5L^*$ sample of galaxies is at least 50\% (2 of 4) and we do not confirm any definitive LyC emitters in the sample. This is consistent with the contamination rates estimated in \citet{nestor11, nestor13}. \citet{nestor13} find significantly lower contamination rates in their faint, Ly$\alpha$-selected sample, suggesting that these types of galaxies are have far higher average escape fractions.  In Section \ref{sect:discussion}, we show that the other $z\sim3$ surveys for LyC are generally consistent with this picture, once varying depth, sample luminosity, and contamination rate are taken into account.  Because the fainter, Ly$\alpha$-selected samples are expected to have higher escape fractions and lower contamination rates, in a subsequent paper, we will present results of additional {\it HST} imaging and Keck spectroscopy of this promising population.

\acknowledgments

{\it Facilities:} \facility{Hubble (WFC3, ACS)}, \facility{Keck: I (LRIS)}, \facility{Keck: II (NIRSPEC)}

Support for program 11636 was provided by NASA through a grant from the Space Telescope Science Institute, which is operated by the Association of Universities for Research in Astronomy, Inc., under NASA contract NAS 5-26555.

The authors wish to recognize and acknowledge the very significant cultural role and reverence that the summit of Mauna Kea has always had within the indigenous Hawaiian community.  We are most fortunate to have the opportunity to conduct observations from this mountain.

\bibliographystyle{apj}
\bibliography{all_ref}

\begin{thebibliography}{62}
\expandafter\ifx\csname natexlab\endcsname\relax\def\natexlab#1{#1}\fi

\bibitem[{{Alavi} {et~al.}(2014){Alavi}, {Siana}, {Richard}, {Stark},
  {Scarlata}, {Teplitz}, {Freeman}, {Dominguez}, {Rafelski}, {Robertson}, \&
  {Kewley}}]{alavi14}
{Alavi}, A. {et~al.} 2014, \apj, 780, 143

\bibitem[{{Bertin} \& {Arnouts}(1996)}]{bertin96}
{Bertin}, E., \& {Arnouts}, S. 1996, \aaps, 117, 393

\bibitem[{{Borthakur} {et~al.}(2014){Borthakur}, {Heckman}, {Leitherer}, \&
  {Overzier}}]{borthakur14}
{Borthakur}, S., {Heckman}, T.~M., {Leitherer}, C., \& {Overzier}, R.~A. 2014,
  Science, 346, 216

\bibitem[{{Boutsia} {et~al.}(2011){Boutsia}, {Grazian}, {Giallongo}, {Fontana},
  {Pentericci}, {Castellano}, {Zamorani}, {Mignoli}, {Vanzella}, {Fiore},
  {Lilly}, {Gallozzi}, {Testa}, {Paris}, \& {Santini}}]{boutsia11}
{Boutsia}, K. {et~al.} 2011, \apj, 736, 41

\bibitem[{{Bridge} {et~al.}(2010){Bridge}, {Teplitz}, {Siana}, {Scarlata},
  {Conselice}, {Ferguson}, {Brown}, {Salvato}, {Rudie}, {de Mello}, {Colbert},
  {Gardner}, {Giavalisco}, \& {Armus}}]{bridge10}
{Bridge}, C.~R. {et~al.} 2010, \apj, 720, 465

\bibitem[{{Casertano} {et~al.}(2000){Casertano}, {de Mello}, {Dickinson},
  {Ferguson}, {Fruchter}, {Gonzalez-Lopezlira}, {Heyer}, {Hook}, {Levay},
  {Lucas}, {Mack}, {Makidon}, {Mutchler}, {Smith}, {Stiavelli}, {Wiggs}, \&
  {Williams}}]{casertano00}
{Casertano}, S. {et~al.} 2000, \aj, 120, 2747

\bibitem[{{Clarke} \& {Oey}(2002)}]{clarke02}
{Clarke}, C., \& {Oey}, M.~S. 2002, \mnras, 337, 1299

\bibitem[{{Conroy} \& {Kratter}(2012)}]{conroy12}
{Conroy}, C., \& {Kratter}, K.~M. 2012, \apj, 755, 123

\bibitem[{{Cowie} {et~al.}(2009){Cowie}, {Barger}, \& {Trouille}}]{cowie09}
{Cowie}, L.~L., {Barger}, A.~J., \& {Trouille}, L. 2009, \apj, 692, 1476

\bibitem[{{Deharveng} {et~al.}(2001){Deharveng}, {Buat}, {Le Brun}, {Milliard},
  {Kunth}, {Shull}, \& {Gry}}]{deharveng01}
{Deharveng}, J.-M., {Buat}, V., {Le Brun}, V., {Milliard}, B., {Kunth}, D.,
  {Shull}, J.~M., \& {Gry}, C. 2001, \aap, 375, 805

\bibitem[{{Dom{\'{\i}}nguez} {et~al.}(2014){Dom{\'{\i}}nguez}, {Siana},
  {Brooks}, {Christensen}, {Bruzual}, {Stark}, \& {Alavi}}]{dominguez14}
{Dom{\'{\i}}nguez}, A., {Siana}, B., {Brooks}, A.~M., {Christensen}, C.~R.,
  {Bruzual}, G., {Stark}, D.~P., \& {Alavi}, A. 2014, ArXiv e-prints

\bibitem[{{Dom{\'{\i}}nguez} {et~al.}(2013){Dom{\'{\i}}nguez}, {Siana},
  {Henry}, {Scarlata}, {Bedregal}, {Malkan}, {Atek}, {Ross}, {Colbert},
  {Teplitz}, {Rafelski}, {McCarthy}, {Bunker}, {Hathi}, {Dressler}, {Martin},
  \& {Masters}}]{dominguez13}
{Dom{\'{\i}}nguez}, A. {et~al.} 2013, \apj, 763, 145

\bibitem[{{Dove} {et~al.}(2000){Dove}, {Shull}, \& {Ferrara}}]{dove00}
{Dove}, J.~B., {Shull}, J.~M., \& {Ferrara}, A. 2000, \apj, 531, 846

\bibitem[{{Eldridge} \& {Stanway}(2009)}]{eldridge09}
{Eldridge}, J.~J., \& {Stanway}, E.~R. 2009, \mnras, 400, 1019

\bibitem[{{Erb} {et~al.}(2006{\natexlab{a}}){Erb}, {Shapley}, {Pettini},
  {Steidel}, {Reddy}, \& {Adelberger}}]{erb06a}
{Erb}, D.~K., {Shapley}, A.~E., {Pettini}, M., {Steidel}, C.~C., {Reddy},
  N.~A., \& {Adelberger}, K.~L. 2006{\natexlab{a}}, \apj, 644, 813

\bibitem[{{Erb} {et~al.}(2003){Erb}, {Shapley}, {Steidel}, {Pettini},
  {Adelberger}, {Hunt}, {Moorwood}, \& {Cuby}}]{erb03}
{Erb}, D.~K., {Shapley}, A.~E., {Steidel}, C.~C., {Pettini}, M., {Adelberger},
  K.~L., {Hunt}, M.~P., {Moorwood}, A.~F.~M., \& {Cuby}, J. 2003, \apj, 591,
  101

\bibitem[{{Erb} {et~al.}(2006{\natexlab{b}}){Erb}, {Steidel}, {Shapley},
  {Pettini}, {Reddy}, \& {Adelberger}}]{erb06b}
{Erb}, D.~K., {Steidel}, C.~C., {Shapley}, A.~E., {Pettini}, M., {Reddy},
  N.~A., \& {Adelberger}, K.~L. 2006{\natexlab{b}}, \apj, 646, 107

\bibitem[{{Finlator} {et~al.}(2012){Finlator}, {Oh}, {{\"O}zel}, \&
  {Dav{\'e}}}]{finlator12}
{Finlator}, K., {Oh}, S.~P., {{\"O}zel}, F., \& {Dav{\'e}}, R. 2012, \mnras,
  427, 2464

\bibitem[{{Fontanot} {et~al.}(2012){Fontanot}, {Cristiani}, \&
  {Vanzella}}]{fontanot12}
{Fontanot}, F., {Cristiani}, S., \& {Vanzella}, E. 2012, \mnras, 425, 1413

\bibitem[{{Fujita} {et~al.}(2003){Fujita}, {Martin}, {Mac Low}, \&
  {Abel}}]{fujita03}
{Fujita}, A., {Martin}, C.~L., {Mac Low}, M.-M., \& {Abel}, T. 2003, \apj, 599,
  50

\bibitem[{{Giallongo} {et~al.}(2002){Giallongo}, {Cristiani}, {D'Odorico}, \&
  {Fontana}}]{giallongo02}
{Giallongo}, E., {Cristiani}, S., {D'Odorico}, S., \& {Fontana}, A. 2002,
  \apjl, 568, L9

\bibitem[{{Glikman} {et~al.}(2011){Glikman}, {Djorgovski}, {Stern}, {Dey},
  {Jannuzi}, \& {Lee}}]{glikman11}
{Glikman}, E., {Djorgovski}, S.~G., {Stern}, D., {Dey}, A., {Jannuzi}, B.~T.,
  \& {Lee}, K.-S. 2011, \apjl, 728, L26

\bibitem[{{Gnedin} {et~al.}(2008){Gnedin}, {Kravtsov}, \& {Chen}}]{gnedin08a}
{Gnedin}, N.~Y., {Kravtsov}, A.~V., \& {Chen}, H.-W. 2008, \apj, 672, 765

\bibitem[{{Grimes} {et~al.}(2009){Grimes}, {Heckman}, {Aloisi}, {Calzetti},
  {Leitherer}, {Martin}, {Meurer}, {Sembach}, \& {Strickland}}]{grimes09}
{Grimes}, J.~P. {et~al.} 2009, \apjs, 181, 272

\bibitem[{{Grimes} {et~al.}(2007){Grimes}, {Heckman}, {Strickland}, {Dixon},
  {Sembach}, {Overzier}, {Hoopes}, {Aloisi}, \& {Ptak}}]{grimes07}
---. 2007, \apj, 668, 891

\bibitem[{{Hurwitz} {et~al.}(1997){Hurwitz}, {Jelinsky}, \&
  {Dixon}}]{hurwitz97}
{Hurwitz}, M., {Jelinsky}, P., \& {Dixon}, W.~V.~D. 1997, \apjl, 481, L31+

\bibitem[{{Inoue}(2010)}]{inoue10}
{Inoue}, A.~K. 2010, \mnras, 401, 1325

\bibitem[{{Inoue} \& {Iwata}(2008)}]{inoue08}
{Inoue}, A.~K., \& {Iwata}, I. 2008, \mnras, 387, 1681

\bibitem[{{Inoue} {et~al.}(2006){Inoue}, {Iwata}, \& {Deharveng}}]{inoue06}
{Inoue}, A.~K., {Iwata}, I., \& {Deharveng}, J.-M. 2006, \mnras, 371, L1

\bibitem[{{Inoue} {et~al.}(2011){Inoue}, {Kousai}, {Iwata}, {Matsuda},
  {Nakamura}, {Horie}, {Hayashino}, {Tapken}, {Akiyama}, {Noll}, {Yamada},
  {Burgarella}, \& {Nakamura}}]{inoue11}
{Inoue}, A.~K. {et~al.} 2011, \mnras, 411, 2336

\bibitem[{{Iwata} {et~al.}(2009){Iwata}, {Inoue}, {Matsuda}, {Furusawa},
  {Hayashino}, {Kousai}, {Akiyama}, {Yamada}, {Burgarella}, \&
  {Deharveng}}]{iwata09}
{Iwata}, I. {et~al.} 2009, \apj, 692, 1287

\bibitem[{{Kelson}(2003)}]{kelson03}
{Kelson}, D.~D. 2003, \pasp, 115, 688

\bibitem[{{Leitet} {et~al.}(2013){Leitet}, {Bergvall}, {Hayes}, {Linn{\'e}}, \&
  {Zackrisson}}]{leitet13}
{Leitet}, E., {Bergvall}, N., {Hayes}, M., {Linn{\'e}}, S., \& {Zackrisson}, E.
  2013, \aap, 553, A106

\bibitem[{{Leitet} {et~al.}(2011){Leitet}, {Bergvall}, {Piskunov}, \&
  {Andersson}}]{leitet11}
{Leitet}, E., {Bergvall}, N., {Piskunov}, N., \& {Andersson}, B.-G. 2011, \aap,
  532, A107

\bibitem[{{Leitherer} {et~al.}(1995){Leitherer}, {Ferguson}, {Heckman}, \&
  {Lowenthal}}]{leitherer95}
{Leitherer}, C., {Ferguson}, H.~C., {Heckman}, T.~M., \& {Lowenthal}, J.~D.
  1995, \apjl, 454, L19+

\bibitem[{{Liu} {et~al.}(2008){Liu}, {Shapley}, {Coil}, {Brinchmann}, \&
  {Ma}}]{liu08}
{Liu}, X., {Shapley}, A.~E., {Coil}, A.~L., {Brinchmann}, J., \& {Ma}, C. 2008,
  \apj, 678, 758

\bibitem[{{Malkan} {et~al.}(2003){Malkan}, {Webb}, \& {Konopacky}}]{malkan03}
{Malkan}, M., {Webb}, W., \& {Konopacky}, Q. 2003, \apj, 598, 878

\bibitem[{{Masters} {et~al.}(2012){Masters}, {Capak}, {Salvato}, {Civano},
  {Mobasher}, {Siana}, {Hasinger}, {Impey}, {Nagao}, {Trump}, {Ikeda}, {Elvis},
  \& {Scoville}}]{masters12}
{Masters}, D. {et~al.} 2012, \apj, 755, 169

\bibitem[{{McLean} {et~al.}(1998){McLean}, {Becklin}, {Bendiksen}, {Brims},
  {Canfield}, {Figer}, {Graham}, {Hare}, {Lacayanga}, {Larkin}, {Larson},
  {Levenson}, {Magnone}, {Teplitz}, \& {Wong}}]{mclean98}
{McLean}, I.~S. {et~al.} 1998, in Society of Photo-Optical Instrumentation
  Engineers (SPIE) Conference Series, Vol. 3354, Society of Photo-Optical
  Instrumentation Engineers (SPIE) Conference Series, ed. {A.~M.~Fowler},
  566--578

\bibitem[{{Mostardi} {et~al.}(2013){Mostardi}, {Shapley}, {Nestor}, {Steidel},
  {Reddy}, \& {Trainor}}]{mostardi13}
{Mostardi}, R.~E., {Shapley}, A.~E., {Nestor}, D.~B., {Steidel}, C.~C.,
  {Reddy}, N.~A., \& {Trainor}, R.~F. 2013, \apj, 779, 65

\bibitem[{{Nestor} {et~al.}(2013){Nestor}, {Shapley}, {Kornei}, {Steidel}, \&
  {Siana}}]{nestor13}
{Nestor}, D.~B., {Shapley}, A.~E., {Kornei}, K.~A., {Steidel}, C.~C., \&
  {Siana}, B. 2013, \apj, 765, 47

\bibitem[{{Nestor} {et~al.}(2011){Nestor}, {Shapley}, {Steidel}, \&
  {Siana}}]{nestor11}
{Nestor}, D.~B., {Shapley}, A.~E., {Steidel}, C.~C., \& {Siana}, B. 2011, \apj,
  736, 18

\bibitem[{{Nonino} {et~al.}(2009){Nonino}, {Dickinson}, {Rosati}, {Grazian},
  {Reddy}, {Cristiani}, {Giavalisco}, {Kuntschner}, {Vanzella}, {Daddi},
  {Fosbury}, \& {Cesarsky}}]{nonino09}
{Nonino}, M. {et~al.} 2009, \apjs, 183, 244

\bibitem[{{O'Meara} {et~al.}(2012){O'Meara}, {Prochaska}, {Worseck}, {Chen}, \&
  {Madau}}]{omeara12}
{O'Meara}, J.~M., {Prochaska}, J.~X., {Worseck}, G., {Chen}, H.-W., \& {Madau},
  P. 2012, ArXiv e-prints

\bibitem[{{Pawlik} {et~al.}(2009){Pawlik}, {Schaye}, \& {van
  Scherpenzeel}}]{pawlik09}
{Pawlik}, A.~H., {Schaye}, J., \& {van Scherpenzeel}, E. 2009, \mnras, 394,
  1812

\bibitem[{{Prochaska} {et~al.}(2009){Prochaska}, {Worseck}, \&
  {O'Meara}}]{prochaska09}
{Prochaska}, J.~X., {Worseck}, G., \& {O'Meara}, J.~M. 2009, \apjl, 705, L113

\bibitem[{{Razoumov} \& {Sommer-Larsen}(2006)}]{razoumov06}
{Razoumov}, A.~O., \& {Sommer-Larsen}, J. 2006, \apjl, 651, L89

\bibitem[{{Shapley} {et~al.}(2005){Shapley}, {Steidel}, {Erb}, {Reddy},
  {Adelberger}, {Pettini}, {Barmby}, \& {Huang}}]{shapley05}
{Shapley}, A.~E., {Steidel}, C.~C., {Erb}, D.~K., {Reddy}, N.~A., {Adelberger},
  K.~L., {Pettini}, M., {Barmby}, P., \& {Huang}, J. 2005, \apj, 626, 698

\bibitem[{{Shapley} {et~al.}(2003){Shapley}, {Steidel}, {Pettini}, \&
  {Adelberger}}]{shapley03}
{Shapley}, A.~E., {Steidel}, C.~C., {Pettini}, M., \& {Adelberger}, K.~L. 2003,
  \apj, 588, 65

\bibitem[{{Shapley} {et~al.}(2006){Shapley}, {Steidel}, {Pettini},
  {Adelberger}, \& {Erb}}]{shapley06}
{Shapley}, A.~E., {Steidel}, C.~C., {Pettini}, M., {Adelberger}, K.~L., \&
  {Erb}, D.~K. 2006, \apj, 651, 688

\bibitem[{{Siana} {et~al.}(2008){Siana}, {Polletta}, {Smith}, {Lonsdale},
  {Gonzalez-Solares}, {Farrah}, {Babbedge}, {Rowan-Robinson}, {Surace},
  {Shupe}, {Fang}, {Franceschini}, \& {Oliver}}]{siana08}
{Siana}, B. {et~al.} 2008, \apj, 675, 49

\bibitem[{{Siana} {et~al.}(2007){Siana}, {Teplitz}, {Colbert}, {Ferguson},
  {Dickinson}, {Brown}, {Conselice}, {de Mello}, {Gardner}, {Giavalisco}, \&
  {Menanteau}}]{siana07}
---. 2007, \apj, 668, 62

\bibitem[{{Siana} {et~al.}(2010){Siana}, {Teplitz}, {Ferguson}, {Brown},
  {Giavalisco}, {Dickinson}, {Chary}, {de Mello}, {Conselice}, {Bridge},
  {Gardner}, {Colbert}, \& {Scarlata}}]{siana10}
---. 2010, \apj, 723, 241

\bibitem[{{Steidel} {et~al.}(1998){Steidel}, {Adelberger}, {Dickinson},
  {Giavalisco}, {Pettini}, \& {Kellogg}}]{steidel98}
{Steidel}, C.~C., {Adelberger}, K.~L., {Dickinson}, M., {Giavalisco}, M.,
  {Pettini}, M., \& {Kellogg}, M. 1998, \apj, 492, 428

\bibitem[{{Steidel} {et~al.}(2003){Steidel}, {Adelberger}, {Shapley},
  {Pettini}, {Dickinson}, \& {Giavalisco}}]{steidel03}
{Steidel}, C.~C., {Adelberger}, K.~L., {Shapley}, A.~E., {Pettini}, M.,
  {Dickinson}, M., \& {Giavalisco}, M. 2003, \apj, 592, 728

\bibitem[{{Steidel} {et~al.}(2010){Steidel}, {Erb}, {Shapley}, {Pettini},
  {Reddy}, {Bogosavljevi{\'c}}, {Rudie}, \& {Rakic}}]{steidel10}
{Steidel}, C.~C., {Erb}, D.~K., {Shapley}, A.~E., {Pettini}, M., {Reddy}, N.,
  {Bogosavljevi{\'c}}, M., {Rudie}, G.~C., \& {Rakic}, O. 2010, \apj, 717, 289

\bibitem[{{Steidel} {et~al.}(2001){Steidel}, {Pettini}, \&
  {Adelberger}}]{steidel01}
{Steidel}, C.~C., {Pettini}, M., \& {Adelberger}, K.~L. 2001, \apj, 546, 665

\bibitem[{{Vanzella} {et~al.}(2010{\natexlab{a}}){Vanzella}, {Giavalisco},
  {Inoue}, {Nonino}, {Fontanot}, {Cristiani}, {Grazian}, {Dickinson}, {Stern},
  {Tozzi}, {Giallongo}, {Ferguson}, {Spinrad}, {Boutsia}, {Fontana}, {Rosati},
  \& {Pentericci}}]{vanzella10b}
{Vanzella}, E. {et~al.} 2010{\natexlab{a}}, \apj, 725, 1011

\bibitem[{{Vanzella} {et~al.}(2012){Vanzella}, {Guo}, {Giavalisco}, {Grazian},
  {Castellano}, {Cristiani}, {Dickinson}, {Fontana}, {Nonino}, {Giallongo},
  {Pentericci}, {Galametz}, {Faber}, {Ferguson}, {Grogin}, {Koekemoer},
  {Newman}, \& {Siana}}]{vanzella12}
---. 2012, \apj, 751, 70

\bibitem[{{Vanzella} {et~al.}(2010{\natexlab{b}}){Vanzella}, {Siana},
  {Cristiani}, \& {Nonino}}]{vanzella10}
{Vanzella}, E., {Siana}, B., {Cristiani}, S., \& {Nonino}, M.
  2010{\natexlab{b}}, \mnras, 404, 1672

\bibitem[{{Whitaker} {et~al.}(2012){Whitaker}, {van Dokkum}, {Brammer}, \&
  {Franx}}]{whitaker12}
{Whitaker}, K.~E., {van Dokkum}, P.~G., {Brammer}, G., \& {Franx}, M. 2012,
  \apjl, 754, L29

\bibitem[{{Willott} {et~al.}(2010){Willott}, {Delorme}, {Reyl{\'e}}, {Albert},
  {Bergeron}, {Crampton}, {Delfosse}, {Forveille}, {Hutchings}, {McLure},
  {Omont}, \& {Schade}}]{willott10}
{Willott}, C.~J. {et~al.} 2010, \aj, 139, 906

\end{thebibliography}

\end{document}